\documentclass[journal,final,10pt]{IEEEtran}
\usepackage[T1]{fontenc}
\usepackage[latin9]{inputenc}
\usepackage{color}
\usepackage{verbatim}
\usepackage{booktabs}
\usepackage{amsmath}
\usepackage{amsthm}
\usepackage{graphicx}

\usepackage{tabu}
\usepackage{amssymb}
\usepackage[font={footnotesize}]{caption}
\usepackage{array, ltablex, multirow,ragged2e}
\usepackage{hyperref}
\usepackage{longtable}
\usepackage{mdframed}

\usepackage{subfigure}

\usepackage{dsfont}

\usepackage{float}
\usepackage[italicdiff]{physics}

\usepackage{algorithm}
\usepackage[noend]{algpseudocode}
\makeatletter
\def\BState{\State\hskip-\ALG@thistlm}
\makeatother

\providecommand{\tabularnewline}{\\}

\definecolor{darkgreen}{RGB}{11,102,35}

\theoremstyle{plain}
\newtheorem{thm}{\protect\theoremname}
\theoremstyle{remark}
\newtheorem{rem}[thm]{\protect\remarkname}
\theoremstyle{plain}
\newtheorem{defn}{Definition}
\theoremstyle{proposition}
\newtheorem{prop}[thm]{Proposition}

\providecommand{\remarkname}{Remark}
\providecommand{\theoremname}{Theorem}
\usepackage[detect-all,detect-weight=true]{siunitx}
\usepackage{xcolor}
\usepackage{multirow}
\usepackage{colortbl}
\usepackage{cite}
\usepackage{array}
\usepackage{url}

\newcommand{\Bcomm}{B_{\rm{c}}}
\newcommand{\Brad}{B_{\rm{r}}}
\newcommand{\Psint}{P_{\rm{s,int}}}
\newcommand{\Pswoint}{P_{\rm{s}}}

\newcommand{\alphaint}{\alpha_{\rm{int}}}

\newcommand{\Ptxcomm}{P_{\rm{c}}}
\newcommand{\Ptxrad}{P_{\rm{r}}}

\newcommand{\deltatcr}{\Delta t_{\rm{C2R}}}
\newcommand{\ttildekcr}{\widetilde{t}_{k,\rm{C2R}}}

\newcommand{\deltatrc}{\Delta t_{\rm{R2C}}}
\newcommand{\ttildekrc}{\widetilde{t}_{k,\rm{R2C}}}

\newcommand{\gammarad}{\gamma_{\rm{r}}}
\newcommand{\gammacom}{\gamma_{\rm{c}}}

\newcommand{\gaintx}{G_{\rm{tx}}}
\newcommand{\gainrx}{G_{\rm{rx}}}

\newcommand{\lambdacom}{\lambda_{\rm{c}}}

\newcommand{\rcomt}{\widetilde{r}_{\rm{c}}(t)}

\newcommand{\sinr}{{\rm{SINR}}}

\newcommand{\sircr}{\text{SIR}_{\text{C2R}}}

\newcommand{\Tksetcr}{ {V}_{k,\rm{C2R}} }
\newcommand{\Tksetrc}{ {V}_{k,\rm{R2C}} }

\newcommand{\textr}{ {\rm{r}} }
\newcommand{\textc}{ {\rm{c}} }

\newcommand{\rrt}{ r_{\textr}(t) }
\newcommand{\rrtkt}{ \widetilde{r}_{\textr}(t, kT) }

\newcommand{\xctkt}{ x_{\textc}(t, kT) }

\newcommand{\fDc}{ f_{D,\textc} }

\newcommand{\fr}{f_{\textr}}
\newcommand{\fcomm}{f_{\textc}}

\DeclareMathOperator\erfc{erfc}
\DeclareSIUnit\decibelSM{dBsm}
\makeatother

\begin{document}

\title{%Potential of Radar Communications in ITS Safety: Interference Mitigation and Networking
RadChat: Spectrum Sharing for Automotive Radar Interference Mitigation}
\author{Canan Aydogdu, Musa Furkan Keskin, Nil Garcia,  Henk Wymeersch and Daniel W. Bliss, \textit{Fellow, IEEE}\thanks{Canan Aydogdu, Musa Furkan Keskin, Nil Garcia,  Henk Wymeersch are with the Department of Electrical Engineering, Chalmers University of Technology, Sweden. e-mail: canan@chalmers.se. 
Daniel W. Bliss is with the Department of Electrical, Computer and Energy Engineering, Arizona State University, USA. This work is supported, in part, by Marie Curie Individual Fellowships (H2020-MSCA-IF-2016) Grant 745706 (GreenLoc)
and a SEED grant from Electrical Engineering Department of Chalmers University of Technology. H.~Wymeersch is supported  by Vinnova grant 2018-01929. }
}
\maketitle

\begin{abstract}
In the automotive sector, both radars and wireless communication are susceptible to interference. However, combining the radar and communication systems, i.e., radio frequency (RF) communications and sensing convergence, has the potential to mitigate interference in both systems. This article analyses the mutual interference of spectrally coexistent frequency modulated continuous wave (FMCW) radar and communication systems in terms of occurrence probability and impact, and introduces RadChat, a distributed networking protocol for mitigation of interference among FMCW based automotive radars, including self-interference, using radar and communication  cooperation. The results show that RadChat can significantly reduce radar mutual interference in single-hop vehicular networks in less than 80 ms. 
%\todo{Furkan: Should we re-write some sentences to emphasize that the article is about R2R interference mitigation (not about joint radar-communications)?}
\end{abstract}

\vspace{-5mm}

\section{Introduction}
Among the main goals of intelligent transportation systems (ITS) are (i) safety: reduce safety threats encountered due to human impact, and (ii) efficiency: provide transportation opportunities in a way that is ecologically and economically sustainable. Two important technological components are automotive radar and vehicular communication, especially for advanced driver assistant systems  and self-driving cars~\cite{patole2017automotive,kong2017}, serving complementary purposes. 

Automotive radar provides \emph{local} situational awareness, giving the vehicle timely and reliable information of the surroundings in the form of radar detections with distance, velocity, angle information. The high localization sensitivity (e.g., up to 3 cm for 76--81 GHz operating radars) and robustness against a variety of conditions (snow/fog/rain~\cite{Ryde2009} or optical effects) of radar, is unfortunately threatened by mutual radar-to-radar (R2R) interference \cite{MosarimFinalReport}. Such interference is expected to be exacerbated with tens of radars deployed on autonomous vehicles in the next decade. Mutual interference results in increased effective noise floor, reduced detection capability and non-existing so-called ghost detections \cite{goppelt2010,goppelt2011analytical, bourdoux2017,brooker2007mutual,MosarimFinalReport}. Techniques for mitigating R2R interference include removing polluted radar waveforms, radar sniffing and avoiding transmission, using frequency diversity and digital beamforming \cite{NHTSAreport}. However, none of these methods guarantees interference-free radar sensing in a cost-efficient and implementable way.

Vehicular communication provides \emph{remote} situational awareness by receiving wireless data packets from other vehicles, even outside the immediate range of local sensors. The communication capabilities built into cars can be divided in cellular, e.g., 4G Long-Term Evolution (LTE); and short range  systems, e.g., WiFi-based 802.11p used in Dedicated Short-Range Communications (DSRC)\cite{kenney2011dedicated} in the USA and ITS-G5 in Europe. While LTE cellular services are suitable for long-term traffic information (e.g., route suggestions), DSRC is specifically dedicated to provide very low latency transmission, critical in communications-based active safety applications, including future collision warning, blind spot warning, braking ahead warning \cite{sjoberg2017cooperative}.  However, DSRC suffers from  communication-to-communication (C2C) interference, especially when many vehicles emit warning messages, in turn affecting system-wide safety. 

%in the limited allocated communication bandwidths, especially due to omnidirectional nature of the transmissions with an increased number of vehicles \cite{rostami2016stability}. C2C interference leads to packet losses, especially in emergency situations when many vehicles emit warning messages, in turn affecting system-wide safety. 
%\begin{figure}[b!]
%\centering{}\includegraphics[width=\linewidth]{figures/fig_RadCom.pdf}
%\caption{Radar Communications can provide a multi-hop radar view in an efficient and fast manner in VANETs}
%\label{fig_RadCom}
%\end{figure}

Finally, some bands allow for dual purpose use: the unlicensed 60 GHz or so-called mmWave band (57--66 GHz) is used for IEEE 802.11ad WiGig communication under restrictions in terms of power emissions, but also for radar\cite{yamawaki199860,kumari2015}. The convergence of radar and wireless communications in mmWave bands can provide benefits for both applications. However, a dual use system must account for four types of interference: not only R2R and C2C, but also communication-to-radar (C2R) and radar-to-communication (R2C) interference. 

%performance by well-designed joint radar and communication systems. Moreover, we believe that a RadCom system, which proves to enhance radar performance, can be used in the current 77GHz radar band.  

In this paper, we propose \emph{RadChat},  a radar and communication cooperation system~\cite{surveyBliss} operating in the 77 GHz radar band,  whose sole purpose is to control and coordinate automotive radar sensors among vehicles via wireless communications in order to mitigate radar interference. 
%\footnote{\textcolor{blue}{RadChat uses communication for the sole purpose of radar interference mitigation (not for any other data transfer). Data communication is outside the scope of this article and therefore left as a future work.}}. 
RadChat is an integrated system with both radar and communication functionality  (i.e., using different waveforms on the same hardware), built with minimal modification from standard frequency modulated continuous-wave (FMCW) based automotive radar, which is the most common, cheap and robust radar format used in the automotive sector today\cite{patole2017automotive}.  Our main contributions are as follows: 
\begin{itemize}
    \item An R2R, C2R, and R2C interference analysis of an FMCW radar and narrowband communication system, which indicates that: (i) R2C and C2R interference impedes reliable communication and radar sensing concurrently. Hence, radar and communication signals with similar powers must not share the same time-frequency resources. (ii) It is possible to ensure negligible R2R interference among different automotive radars if the FMCW radar chirp sequences start sweeping the frequency band at different time slots. 
    \item A protocol for the physical (PHY) and medium access (MAC) layers, which is able to essentially reduce R2R radar interference  while meeting automotive radar sensing requirements (i) among both radars on different vehicles and radars mounted on the same vehicle, (ii) in a fairly short time (80 ms). 
    \item An in-depth analysis of the performance of RadChat, compared to standard FMCW in a single-hop dense vehicular ad-hoc network (VANET). 
\end{itemize}
 The additional capabilities of RadChat to enable inter-vehicle communication is outside the scope of this article and therefore left as a future work.

%RadChat is an FMCW-based radar communications framework, which implements the physical (PHY) and data link control (DLC) layers. To the best of our knowledge, this is the first study proposing higher layer solutions for the FMCW-based RadCom. RadChat has the potential to (i) eliminate R2R radar interference, (ii) decrease C2C interference by the directivity of radar antennas and enable low-latency link communications, (iii) increase hardware and spectrum efficiency. Moreover, RadChat offers a low-latency networking solution, where the map of radar-sensed surrounding objects are immediately shared via wireless communications, providing a multi-hop radar view of an extended area as illustrated with an example scenario in Fig.\ref{fig_RadCom}. The focus of this article is the implementation of RadChat in single-hop VANET setting, where multi-hop implementation is considered as a future work.
 
%The coexistence of radar and communications of RadCom units, whilst eliminating R2R interference and decreasing C2C interference, introduces two other types of interference: (i) radar-to-communication (R2C) interference, where a communication link is affected by a radar signal; (ii) communication-to-radar (C2R) interference, where a communication link causes interference to radar. The second contribution of this article is the analysis of these interference types. 

The remainder of this paper is organized as follows. After a brief literature review in Section \ref{sec_back}, an FMCW-based {radar and communication cooperation} system model is introduced in Section \ref{sec_theory}, which is followed by a detailed analysis of R2R, C2R and R2C interference in Section \ref{sec_int}. The RadChat framework is described in Section \ref{sec_model} including MAC and PHY layers. The  R2R interference mitigation and networking performance of RadChat is investigated and results are presented in Section \ref{sec_results}.  

\section{Related Work}
\label{sec_back}

\subsection{Classification of Joint Radar and Communications Systems}
 Joint radar and communication systems can be grouped in three categories: coexistence, cooperation and co-design. Coexistence aims to mitigate inter-system interference without information exchange. In cooperation, information is explicitly shared among both systems to mitigate interference~\cite{surveyBliss}. Finally, co-design methods require both systems to be designed jointly from the ground up, not necessarily using the same hardware~\cite{surveyBliss}, but generally using the same waveform\cite{smartBliss}. 

%\textcolor{blue}{Joint radar-communications systems can be grouped in three categories: coexistence, cooperation and co-design. In \textit{co-existence} scenarios, radar and communications transceivers share the same time-frequency-space resource and thus induce interference onto one another, which, in most cases, must be handled without information exchange between the two systems \cite{surveyBliss}. On the other hand, \textit{cooperative} radar-communications rely on information sharing for joint mitigation of mutual interference and performance improvement of both systems \cite{surveyBliss}. Finally, for the \textit{co-design} approach, the transceivers have the dual-functional capability of radar sensing and communications via a single waveform and can be designed from top to bottom to accomplish the two functionalities simultaneously \cite{RadCom_2017}. In this study, we first investigate a \textit{co-existence} scenario between FMCW radars and narrowband communication systems through mutual interference analysis (including R2R, R2C and C2R), and, then, propose a \textit{cooperative} radar-communications approach for coordination of FMCW radar transmissions via a spectrally isolated communications channel.}
%%%%%%%%%%%%%%%%%%%%%%%%%%%%%%%%%%%%%%%%%%%%%

\subsubsection{Coexistence}
Radar communication coexistence was shown to increase the efficiency of the underutilized radar spectrum and solve the spectrum scarcity of communication systems~\cite{survey185,survey186}, and was therefore used in many different ways~\cite{surveyBliss,survey2016,survey2017,survey2018} with the aim of sharing the same frequency band without radar and communication interfering each other.  Different from these studies, our primary goal is not the spectral efficiency but mitigation of R2R interference, which turns out to be a problem in VANETs. We target to achieve this goal by help of communications,  using the same hardware but different waveforms, which  falls into the scope of radar and communication cooperation, described next.
\subsubsection{Cooperation}
For vehicular applications, the combination of communication and radar  has been considered in various forms \cite{takeda1998,zhong2017,sturm2011}. Estimation and information theoretic approaches were conducted on pulsed radars~\cite{Bliss_Inner_Bounds_TSP} and FMCW radars~\cite{blissFMCW} in the joint multiple-access channels. Radar and communication  cooperation, using the same hardware  but different waveforms for radar and communications, was used in the 79 GHz band with the goal of improving individual pulsed radar sensing accuracy through communications, where the radar and communications use the channel in a time-division-medium-access (TDMA) manner controlled by a central unit~\cite{survey171}.

\subsubsection{Co-design}
 There are several radar communication co-design methods proposed in the literature. IEEE 802.11ad preamble is used as a radar signal for a vehicular environment in the 60 GHz band~\cite{kumari2015}, a monopulse radar with frequency-shift-keying is used to incorporate  communication data in a time-division-multiplexed (TDM) manner~\cite{FSK2007} and minimum shift keyed linear frequency modulated spread spectrum signals (MSK-LFM/SS) are used\cite{zhong2017}. Orthogonal frequency division multiplexing (OFDM) has been the most extensively investigated option for radar communications co-design\cite{donnet2006,garmatyuk2011,sturm2011,kumari2015,junil2016,eylem} due to its high degree of flexibility and high performance under different propagation conditions~\cite{nuss2007,falcone2010,reichardt2012,yoke2016}. However, due to the cost-efficient, low-rate analog-to-digital convertors (ADC) preferred in automotive radars today, OFDM cannot fully occupy the radar band (77--81 GHz), limiting its distance resolution capability.

\subsection{Medium Access Control for Cooperation}
Most studies on cooperation between radar and communication have focuses on the physical layer ~\cite{survey2016}.  In our prior work \cite{CananRadConf,CananPimrc}, we have shown the potential of higher-layer coordination of automotive radars through communications for decreased R2R interference. % for two facing radars, whereas the same system was shown to increase the probability of detection of vulnerable road users by eliminating ghost targets\cite{CananPimrc}. This article complements these two previous studies by a comprehensive explanation of the method in large scale networks by the added functionality of self-interference mitigation among radars mounted on the same vehicle. 
There are  few other studies including the higher layers in radar and communication cooperation \cite{eylemMAC, unifiedMAC}. A separate dedicated radio is used for communication control in addition to a radar communications unit employing OFDM for communications in~\cite{eylemMAC} with an emphasis on data communications rather than interference mitigation. Another MAC approach employing time division multiplexing among radar and communications is introduced in~\cite{unifiedMAC}, where a preamble is added just prior to the radar. As a result, the radar is treated as a packet in CSMA-based communications and radar sensing has no priority over communications, which might end up with low radar sensing duty cycle in case of radar congestion. %\textcolor{blue}{In this article, a radar communications cooperation method is proposed with the goal of mitigation of interference while meeting automotive radar sensing requirements.} 

%We previously used FMCW radar as a means for \textcolor{blue}{cooperation based radar communications} %, which is widespreadly used in the automotive industry due to its robustness, simplicity and low-cost \cite{CananRadConf,CananPimrc}. 

%It was shown that consideration of higher layers improved both radar and communication capabilities compared to coexistence studies focusing solely on the physical layer~\cite{survey2016}. The analysis of the R2R interference conducted in \cite{CananRadConf}, showed that higher-layer coordination of automotive radars through communications decreased R2R interference for two facing radars, whereas the same system was shown to increase the probability of detection of vulnerable road users by eliminating ghost targets\cite{CananPimrc}. This article complements these two previous studies by a comprehensive explanation of the method in large scale networks by the added functionality of self-interference mitigation among radars mounted on the same vehicle. 

%The pedestrian detection probability is shown to increase with the proposed simple RadCom scheme in~\cite{CananPimrc} due to mitigation of R2R interference. This article, investigates the C2R and R2C interference emerging due to coexistence of radar and communications and introduces the proposed RadChat protocol. 

\section{System Model}
\label{sec_theory}
\begin{figure}
\centering{}\includegraphics[clip,width=0.9\columnwidth]{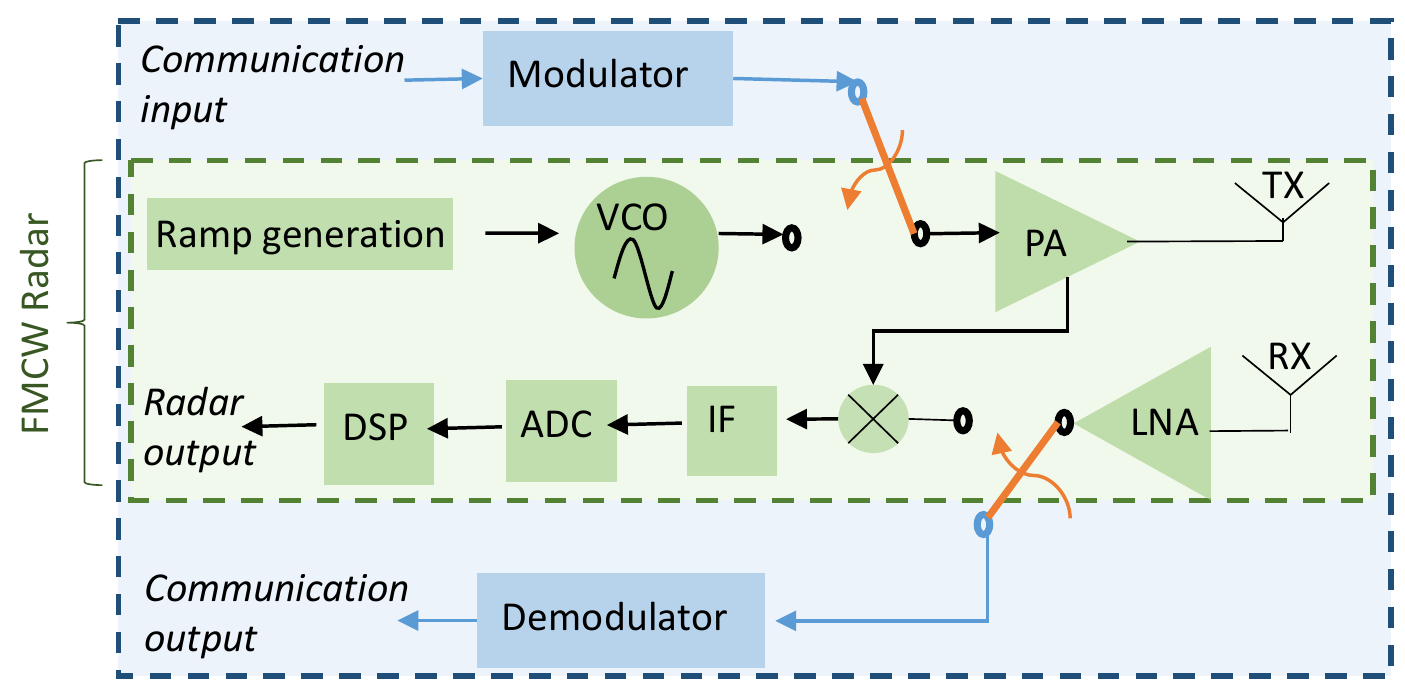}
\caption{Illustration of the hardware of a RadChat unit. The unit reuses most of the hardware for communication (blue) and radar (green) functionalities.}
\label{fig_RadComHardware}
\end{figure}

\subsection{RadChat Unit}
A {RadChat unit} is a modified automotive FMCW radar hardware, where the input to the conventional FMCW radar transmitter is switched between radar and communication and likewise the receiver antenna output is switched between the radar and communication receiver module as illustrated in Fig.~\ref{fig_RadComHardware}. We assume a  homogeneous VANET with identical RadChat units, where all vehicles have the same radar and communication parameters (radar frame time, radar slope, radar bandwidth/carrier frequency, communication bandwidth/carrier frequency/modulation scheme, etc). RadChat units transmit and receive either radar or communication signals, but not both radar and communication signals simultaneously. The communication input/output and radar output of the RadChat unit is connected to the MAC layer of RadChat, introduced in Sec.~\ref{sec_model} where spectrum sharing among radar and communications as well as among different RadChat units is presented. We now describe the operation of the RadChat unit. 
\begin{figure}
\centering{}\includegraphics[width=\linewidth]{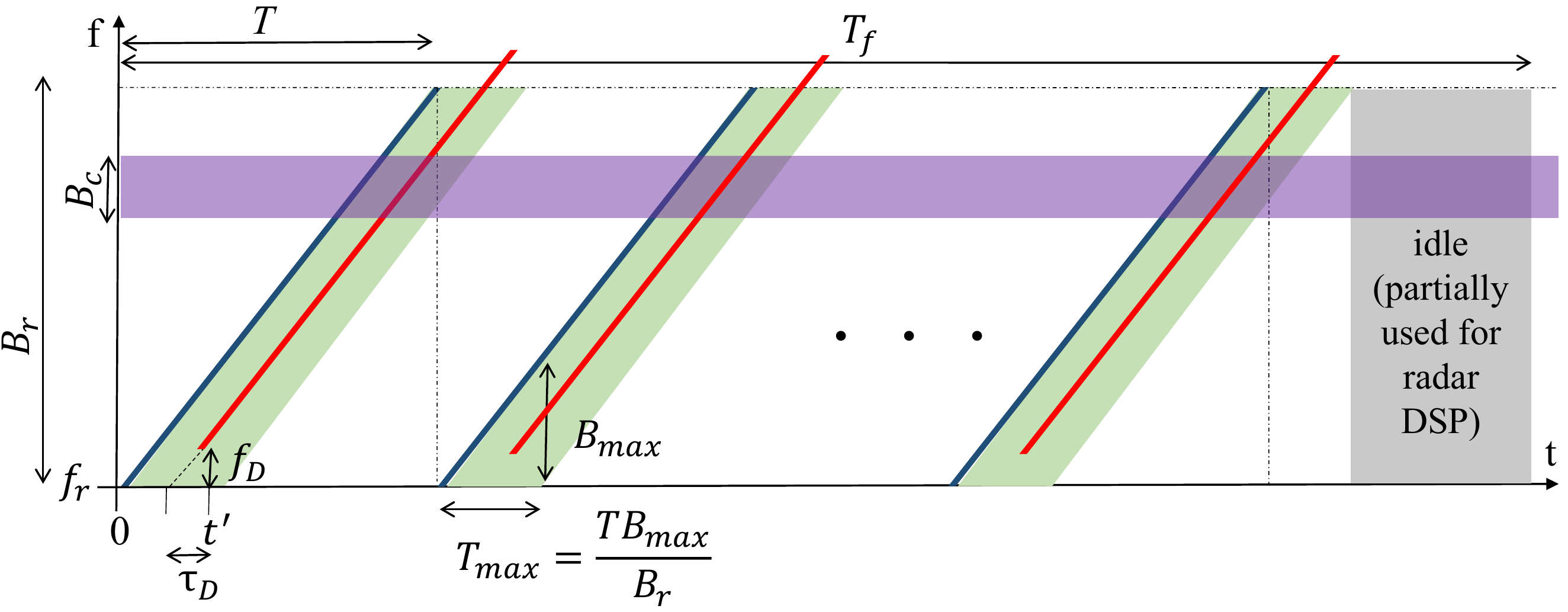}
\caption{FMCW sawtooth radar waveform occupying $B_\text{r}$ bandwidth with simultaneous communication occupying $\Bcomm$ bandwidth. The transmitted radar chirp sequence (blue lines) are received (red lines) with a Doppler frequency shift of $f_D$. The radar receiver is tuned to process radar reflections arriving inside the green band, which corresponds to the bandwidth of interest $B_\text{max}$.}
\label{fig_chirpJournal}
\end{figure}

\subsection{RadChat Transmitter}
The transmitter of the RadChat unit either transmits radar signals or communication signals, but not both at the same time as illustrated in Fig.\ref{fig_RadComHardware}.

\subsubsection{Radar Transmitter}
We consider a sequence of frequency modulated continuous waves, i.e., chirps, transmitted by an FMCW radar,
\begin{equation}
s_{\text{r}}(t)=\sqrt{\Ptxrad}\sum_{k=0}^{N-1}c(t-kT)\label{eq:TXwaveform}
\end{equation}
where $c(t)$ is a chirp of the form \cite{Linear_FMCW_1992,patole2017automotive}
\begin{equation}
c(t)= \begin{cases}
e^{j \phi(t)} \,, &~ 0 \leq t \leq T  \\
0\,, &~ \text{otherwise}
\end{cases} \label{eq:chirp}
\end{equation}
with $\phi(t) = 2\pi\left(\fr+\frac{B_\text{r}}{2T}t\right)t $, where $\Ptxrad$ is the radar transmit power, $B_\text{r}$ denotes the
radar bandwidth (typically 1\textendash4 GHz), $\fr$ is the carrier frequency (around 77 GHz), $T$ is the chirp duration, and $N$ is the number of chirps per frame. The frame time $T_{f}=NT + T_{\text{idle}}$ comprises $NT$ plus the idle and processing time. The instantaneous frequency of $c(t)$ is given by $f(t) = \frac{1}{2 \pi} \dv{\phi(t)}{t} = \fr + \frac{B_\text{r}}{T} t$. Fig.~\ref{fig_chirpJournal} illustrates a typical FMCW sawtooth radar waveform with a chirp sequence starting at $(t=0,f=f_\text{r})$ in time-frequency domain. The reflected chirp sequences are received starting at $(t',f_\text{r}+f_D)$, due to a round trip delay of $t'$ and a Doppler frequency shift of $f_D$. The green band corresponds to the bandwidth $B_{\text{max}}$, which is defined as the bandwidth of interest at the radar receiver. Note that $B_{\text{max}} \leq f_s$, where $f_s$ is the ADC bandwidth, assuming a complex baseband radar architecture \cite{TI_Complex_Radar}. The radar receiver filters out the radar reflections arriving with frequencies outside $(\fr+\frac{B_\text{r}}{T}t\big)+[-B_{\text{max}},0]$. $B_\text{max}$ is proportional to the maximum delay of radar reflections taken into account ($T_\text{max}$) and the maximum detectable range ($d_\text{max}$). %The grey band  in Fig.~\ref{fig_chirpJournal} is the idle time of a radar, which is partially used for radar signal processing. The region outside of the green stripes, constitute a large fraction of unused time-frequency resources for a classical FMCW radar. The RadCom unit proposed in this article switches to communication mode in this grey region.

\begin{rem}
The FMCW radar waveform parameters, such as $B_\text{r}$, $T$, $T_f$ and $N$, are set to meet requirements  on the maximum detectable range ($d_\text{max}$) and  maximum detectable relative velocity ($v_\text{max}$), as well as range and velocity resolution. 
\end{rem}
\subsubsection{Communication Transmitter}
During communication mode, the transmitted bandpass signal is \cite{goldsmith2005wireless}
\begin{equation}\label{eq_comm_signal}
    s_{\text{c}}(t) = \sqrt{\Ptxcomm} x(t)  e^{j  2\pi  \fcomm  t}
\end{equation}
where $x(t)$ represents the complex baseband signal with a bandwidth of $\Bcomm$ after pulse shaping\footnote{The sampling rate satisfies $\Bcomm \leq f_\text{s}$, which is generally on the order of $10-50$ MHz for automotive radars.} and $\fcomm$ is the communication carrier frequency. We consider a communication signal 
\begin{align}\label{eq_comm_signal_baseband}
    x(t) = \sum_{k=0}^{N_{\text{c}}-1} a_k p(t-k(1+\alpha)/\Bcomm), 
\end{align}
where $p(t)$ is a unit-energy pulse of bandwidth $\Bcomm$ with roll-off $\alpha \ge 0$, $a_k$ are unit-energy transmit data symbol. 

\subsection{RadChat Receiver}
The receiver of the RadChat unit either receives radar signals or communication signals, not both at the same time.
\subsubsection{Radar Receiver}\label{sec_radar_rec}
Considering a single target at distance $d$, the received back-scattered bandpass radar signal (at the co-located receiver) is
\begin{equation}
\rrt = \gammarad \sum_{k=0}^{N-1} c\left( t - 2d/c - kT \right) + w(t) \label{eq_radar_received}
\end{equation}
%\begin{align}
%r_{\text{r}}(t) = & {\gamma_r }\sum_{k=0}^{N-1}  e^{j2\pi\big(\fr-2f_D+\frac{\Brad}{2T}(t-kT-\frac{2d}{c})\big)(t-kT-\frac{2d}{c})}+w(t) \label{eq_radar_received}
%\end{align}
where a Doppler shift will be observed due to a time-varying distance $d = d_0 - v t$ with relative radial velocity $v$ and initial distance $d_0$,
\begin{equation}\label{eq_radar_received_power}
    \gammarad = \sqrt{\Ptxrad \gaintx \gainrx \sigma \lambda_\text{r}^2/((4\pi)^3 d^4)}
\end{equation}
for target radar cross section\footnote{Throughout the paper, RCS is assumed to be non-fluctuating over the frame time $T_f$.} (RCS) $\sigma$, transmitter and receiver antenna gains $G_{\text{tx}}$ and $G_{\text{rx}}$, $w(t)$ is additive white Gaussian noise (AWGN) with power spectral density $N_{0}$, $c$ denotes the speed of light and $\lambda_\text{r}$ is the wavelength of the radar carrier. 
% between vehicles (note that a positive $v$ corresponds to approaching vehicles and a positive Doppler shift, which leads to a decreased frequency difference between the transmitted and reflected radar signal). 
The received signal is processed by the following blocks \cite{refOfMatlab}:
a mixer, an ADC, and a digital processor (Fig.~\ref{fig_RadComHardware}). Then, after processing the bandpass signal in \eqref{eq_radar_received} through the receive chain, the 
%The mixer multiplies the conjugate of the received signal with a copy of the transmitted chirp \cite{Linear_FMCW_1992,FMCW_SAR_2007,patole2017automotive}. After low-pass filtering the resulting intermediate frequency (IF) signal, the mixer will output a signal with multiple harmonics at frequencies proportional to the time difference between the transmitted chirp and the received chirps. The output of the mixer is then sampled by the ADC, with sampling interval $T_{s}$, and passed to the digital processor which will detect and estimate the frequencies. The ADC bandwidth $1/(2 T_{s})$ is generally on the order of 10--50 MHz and is
%thus much smaller than $\Brad$. 
sampled baseband ADC output is as following for the chirp $k$, $t=n/f_\text{s}$, $n=0,\ldots,\lfloor Tf_\text{s} \rfloor $
\begin{align}\nonumber
\rrtkt &= \gammarad  e^{  j2\pi  t \left( -\frac{2 d_0}{c} \frac{\Brad}{T} + 2 \frac{v}{c} \fr - 2 \frac{v}{c} \frac{\Brad}{T} t \right) } \\ \label{eq_rx_waveform_general}  & ~~~~ \times e^{  j2\pi  k T \left(2 \frac{v}{c} \fr - 2 \frac{v}{c} \frac{\Brad}{T} t \right) - \frac{2 d_0}{c} \fr   } + w(t, kT)
\end{align}
%\begin{equation}\label{eq_rx_waveform_general}
 %   \rrtkt = \gamma_\text{r}  e^{  j2\pi \left[ t \left( -\frac{2 d_0}{c} \frac{B_\text{r}}{T} + 2 f_D \right) + 2 f_D k T - \frac{2 {d_0}}{c} \fr \right]  } + w(t, kT)
%\end{equation}
%\begin{equation}\label{eq_rx_waveform_general}
  %  \rnkbar = {\gamma_r}e^{j2\pi\frac{n}{f_s} (2f_D-\frac{\Brad}{2T}\frac{2d}{c})} e^{j2\pi \fr\frac{2d}{c}} + \wnk
%\end{equation}
%with
%\begin{align}
%\rnk = & \sqrt{\gamma \Ptx %d^{-4}}e^{j2\pi\big(f_{c}-2f_D+\frac{\Brad}{2T}(nT_s-kT-\frac{2d}{c})\big)(nT_s-kT-\frac{2d}{c})}
%\label{eq_RXwaveform}
%\end{align}
where $0 \leq t \leq T$ denotes the time from the beginning of the $k$th chirp. Assume that we have a narrowband signal, i.e., $\Brad \ll \fr$, and that target displacement during a chirp is much smaller than the wavelength, i.e., $vT \ll \lambda_\text{r}$. %\footnote{Note that these two assumptions are generally satisfied for typical FMCW radar parameters. For instance, for $\fr = 77\,$GHz, $\Brad = 1\,$GHz, $d_0 = 50\,$m, $v=40\,$m/s and $T = 20\,\mu$s, we have $\frac{\fr}{\Brad} = 77$ and $\frac{d_0}{v T} = 62500$.}. %The Doppler frequency shift is $f_D={v\fr}/{c}$, assuming that $B_\text{r} \ll f_\text{r}$ so that the Doppler shift is evaluated to be independent over time.\NIL{NIL: There is an important implicit approximation here that is not mentioned.
%In a narrow-band signal, where the carrier frequency is much larger than the bandwidth, this formula would be correct. For this formula to be valid we should prove by a simple calculation that the bandwidth divided by carrier frequency is indeed very small.} \CA{QUESTION to NIL: Do you think the previous sentence answers your concern Nil?} 
Based on these assumptions, the signal in \eqref{eq_rx_waveform_general} can be approximated as \cite{patole2017automotive}
\begin{align}\label{eq_rec_signal_2}
\rrtkt &= \gammarad  e^{ j2\pi \left[  -\frac{2 d_0}{c} \frac{\Brad}{T} t + 2 f_D k T - \frac{2 d_0}{c} \fr \right]  } + w(t, kT) 
\end{align}
where $f_D={v\fr}/{c}$ is the Doppler shift. A common approach for range-Doppler retrieval in FMCW radar is to compute the fast Fourier transform (FFT) of the signal in \eqref{eq_rec_signal_2} over both fast time $t$ and slow time $k$ (with windowing functions \cite[Ch.~5.3.1]{richards2005fundamentals}), yielding peaks at frequencies corresponding to $d_0$ and $f_D$, respectively, and detect the peaks in the range-Doppler domain. 

%Note that for typical FMCW radar parameters, $\frac{2 d_0}{c} \frac{B_\text{r}}{T}$ is on the order of $10\,$MHz, while $2 f_D$ is on the order of $10\,$kHz, leading to negligible range-Doppler coupling in fast-time range estimation in \eqref{eq_rx_waveform_general}.

\subsubsection{Communication Receiver}
During communication mode, the complex baseband received signal is \cite{goldsmith2005wireless}
\begin{equation}
    \rcomt = \gammacom x(t - {d}/{c})e^{j 2 \pi (\fDc t - \fcomm d_0/c ) }+w(t) % \exp \left( j 2 \pi \left[ \frac{\fcomm v}{c} t - \frac{\fcomm d_0}{c} \right] \right)
\end{equation}
where
\begin{equation}\label{eq_gammacom}
    \gammacom = \sqrt{\Ptxcomm \gaintx \gainrx \lambdacom^2/(4\pi d)^2} 
\end{equation}
under the assumption of free-space propagation environment \cite{goldsmith2005wireless,molisch2012wireless}, $\fDc = v \fcomm /c$ and $\lambdacom$ is the wavelength of the communications carrier.
%$d = d_0 - vt$ is the distance between vehicles (with $d_0$ and $v$ representing the initial distance and the relative velocity, respectively) and $\lambdacom$ is the wavelength of the communications carrier.

%the sampled back-scatter signal, sample $n$ of
%chirp $k$ is of the form \cite{patole2017automotive}
%\begin{equation}\label{eq_rx_waveform_general}
%    \rnk = \rnkbar + \wnk
%\end{equation}
%with
%\begin{align}\nonumber
%\rnkbar = & \sqrt{\gamma \Ptx d^{-4}}\exp\left(j2\pi\frac{B}{T} \big[(2d/c -2\tau_{D})nT_{s} - \textcolor{red}{2\tau_{D}kT} \big] \right) \\ & \textcolor{red}{\times \exp\left( j2\pi \frac{2d}{c} f_c \right)}   \label{eq:RXwaveform}
%\end{align}
%where $\gamma= \gaintx \gainrx \sigma \lambda^2/(4\pi)^3$, for  target radar cross section (RCS) $\sigma$, transmitter and receiver antenna gains $G_{\text{tx}}$ and $G_{\text{rx}}$, Doppler time shift $\tau_{D}={Tvf_c}/{(Bc)}$, in which $c$ denotes the speed of light, $v$ is the relative velocity between vehicles (note that a positive $v$ corresponds to approaching vehicles and a positive Doppler shift, which leads to a decreased time difference between the transmitted and reflected radar signal), $w_{n}^{(k)}$ is additive white Gaussian noise (AWGN) with variance $N_{0}$.  A common approach to \textcolor{red}{range-Doppler} retrieval in FMCW radar is to compute the fast Fourier transform (FFT) of the signal \textcolor{red}{$r_{n}^{(k)}$ over both fast time dimension $n$ and slow time dimension $k$}, and detect the peaks in the \textcolor{red}{range-Doppler} domain.

\section{Interference Analysis}
\label{sec_int}
\begin{figure}[t!] \centering
\subfigure[]{\includegraphics[width=0.9\linewidth]{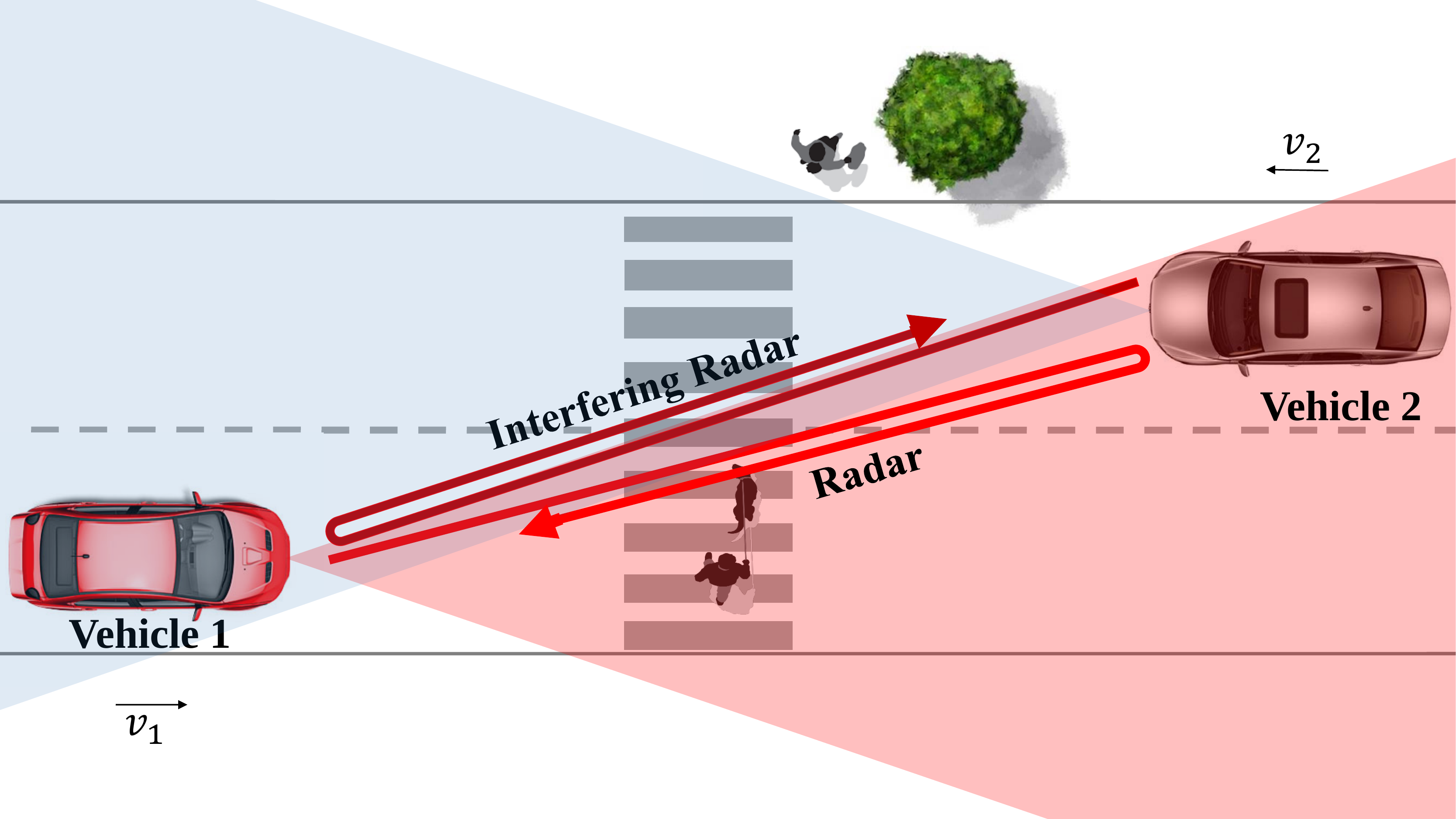}
\label{fig_scenarioIntR2R}}
\subfigure[]{\includegraphics[width=0.9\linewidth]{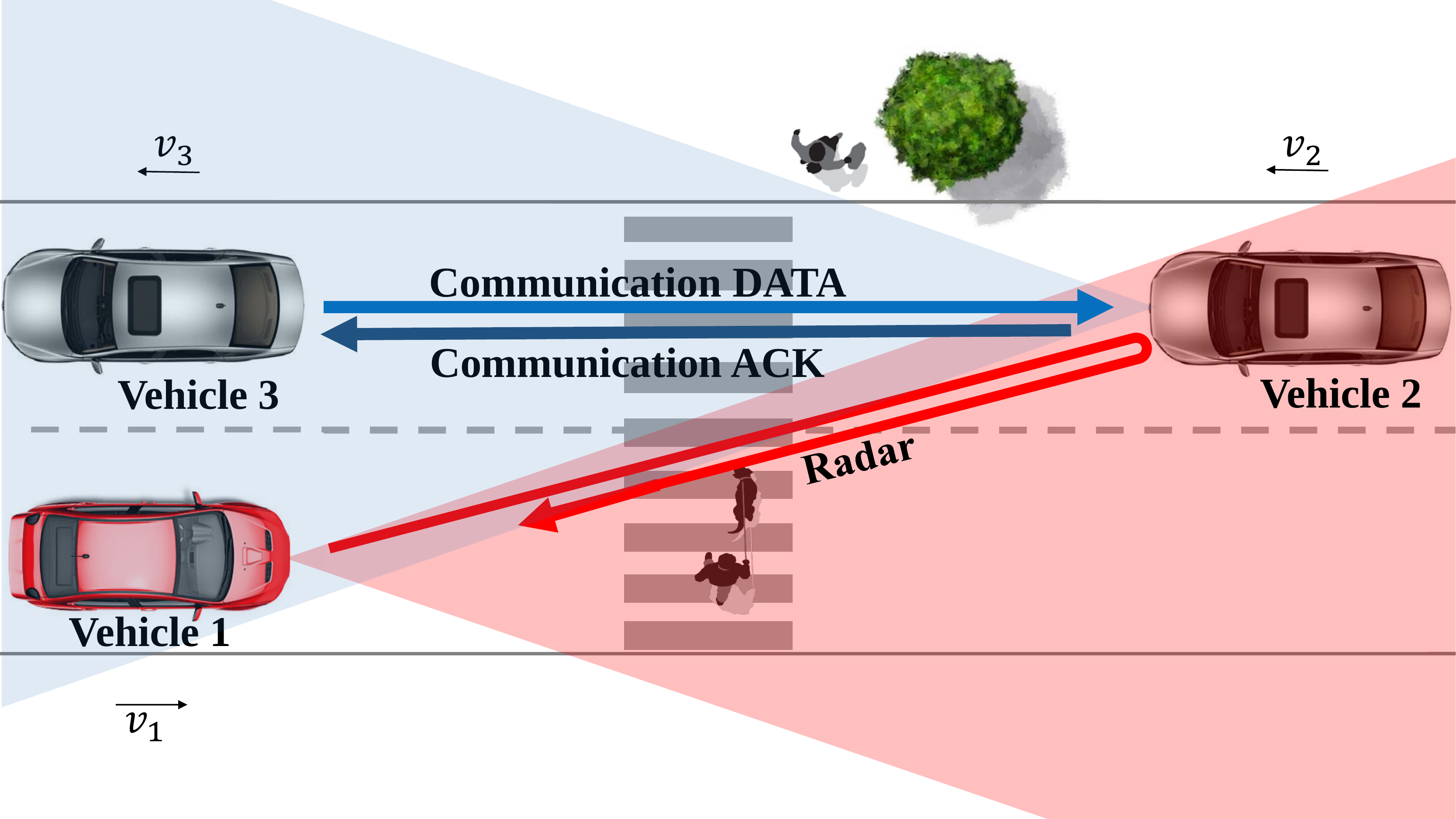}
 \label{fig_scenarioInt} }
 \caption{Scenarios for investigation of  a) R2R interference, b) C2R interference experienced at Vehicle 1 and R2C interference at Vehicle 2.}
 \label{fig_scenario}
\end{figure}

To gain understanding in the three types of interference (R2R, C2R and R2C), we consider two simple scenarios given in Fig.~\ref{fig_scenarioIntR2R}--\ref{fig_scenarioInt}. 
For the R2R interference,  the scenario given in Fig.~\ref{fig_scenarioIntR2R} demonstrates a case where two front-end radars of two vehicles illuminate each other's field of view (FoV). For the C2R and R2C case, we consider the three-vehicle scenario given in Fig.~\ref{fig_scenarioInt}, where the rear-end radar of Vehicle 3 sends communication data to  Vehicle 2 and receives acknowledgements\footnote{Other data packets may follow the ACK from Vehicle 2 to Vehicle 3, which cause C2R interference at Vehicle 1. We assume that the communication signal occupies the channel during the whole FMCW chirp sequence independent of the packet type.} (ACK), while the front-end radar of Vehicle 1 is simultaneously performing radar sensing. This ends up with C2R interference at Vehicle 1, where the ACKs of communication disturb the radar signal of Vehicle 1. R2C interference is analyzed for the same scenario at Vehicle 2, where the communication data is affected by the radar of Vehicle 1. 

We assume that all FMCW waveform parameters (frame time, chirp duration, slope, bandwidth, number of chirps per frame, duty cycle, etc.) are the same for all radars for the sake of simplicity. In all analyses, signals are assumed to be in the main beam of the respective antennas so that the antenna gains are taken equal. Table \ref{tab:InterferenceTable} provides a concise comparison of the different types of interference, which will be derived in the subsequent sections. 
%We assume Swerling~$0$ type RCS for radar targets \cite{richards2005fundamentals}. \NIL{Reviewers may complaint because Swerling models were designed for point targets, like a plane in the very far field.
%Also, we need to motivate why the choice of Swerling 0 even if the reason is because "it's easier".}\CA{Comment to NIL: As we talked, you are gonna put here something about your own comment.}

\begin{table}%[h]
\centering
\begin{tabular}{|c|c|c|}
\hline 
\textbf{Type} &  \textbf{SIR} & \textbf{Probability}\tabularnewline
\hline 
\hline 
R2R &   $\mathcal{O}(\sigma/d^{2})$ & $4UB_{\text{max}}/B_{\text{r}}$\tabularnewline
\hline 
C2R &   $\mathcal{O}(\Ptxrad\sigma/(\Ptxcomm d^{2}))$ & $ U \min\{ B_\text{max} + \Bcomm, B_\text{r} \} /  B_\text{r}$\tabularnewline
\hline 
R2C &   $\mathcal{O}(\Ptxcomm/\Ptxrad)$ & $U \min\{  \Bcomm, B_\text{r} \} /  B_\text{r} $\tabularnewline
\hline 
\end{tabular}
\caption{Signal-to-interference ratio and probability of interference values for the different interference cases, where radar target and interference sources are located at the same distance $d$, and $U$ is the radar duty cycle.}
%Properties of different interference cases with radar target and interference source are at the same distance $d$, where $U$ is the radar duty cycle.}
\label{tab:InterferenceTable}
\end{table}

\subsection{R2R Interference Analysis}
\label{subsec:R2R}

R2R interference might occur in two different ways, both of which are considered in this article: (i) direct line-of-sight (LoS) interference, (ii) bistatic radar returns or reflected interference, when either a victim vehicle receives a reflected interfering radar signal from another vehicle or one RadChat unit at the victim vehicle receives a reflected radar return of another RadChat unit on the same victim vehicle.

\subsubsection{Impact and Power of R2R Interference}\label{sec_imp_r2r} 
R2R interference affects radar performance in a number of ways: it leads to an increase of the effective noise floor or false alarms (ghost targets), which are apparent targets with high intensity that are not actually present. In our system where all radars have the same parameters, ghost targets will be the dominant effect, while effective noise floor increase occurs when radars have different chirp parameters~\cite{Kim2018}. 

\begin{figure}[t!]
\vspace{-0.2in}
\centering
\includegraphics[width=1\columnwidth]{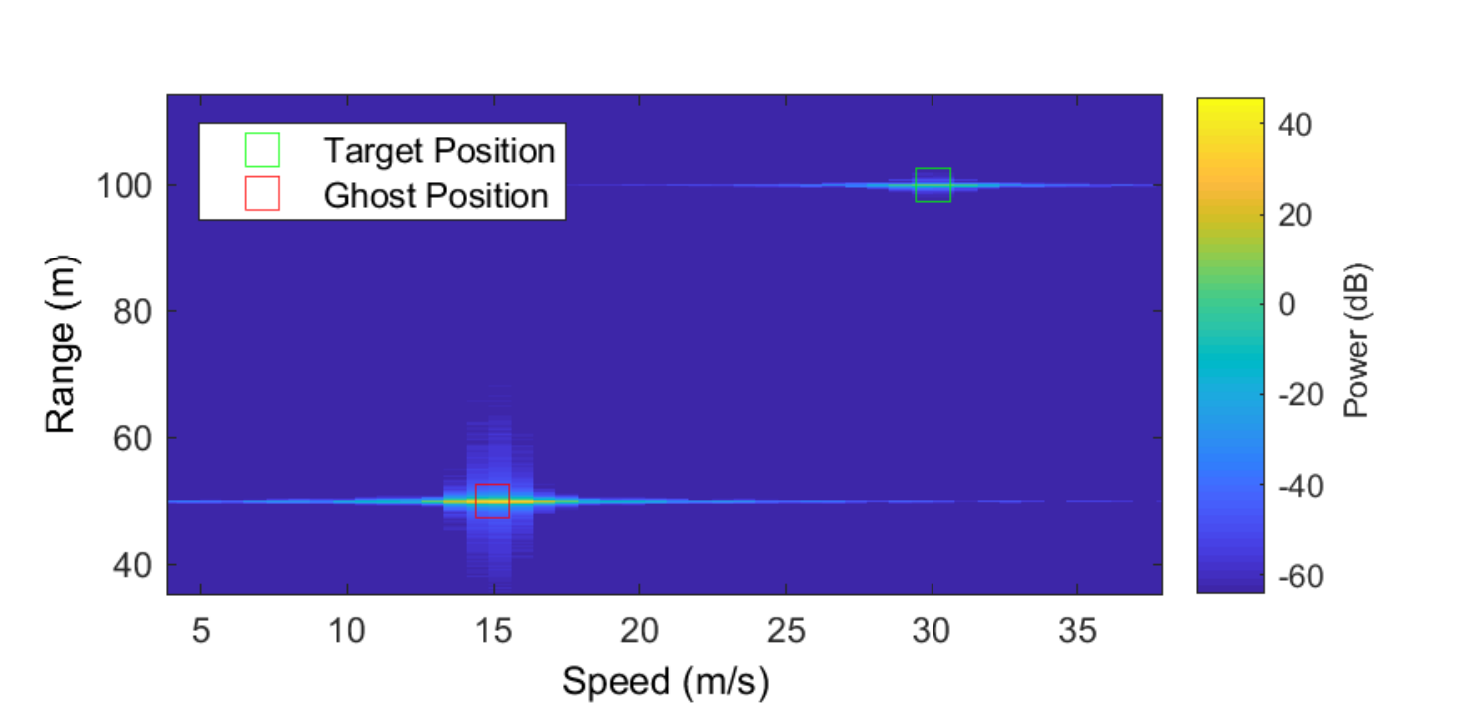}
\caption{Radar range-Doppler map in the presence of R2R interference, where $d = 100\,$m, $v=30\,$m/s and $\Ptxrad = 5\,$mW.}
\label{fig_r2r_ser_vs_bw}
\vspace{-0.2in}
\end{figure}

\textit{Example:}
The range-Doppler plot illustrating the R2R interference for two vehicles approaching with $v=30$ m/s relative speed at $d=100$ m is given in Fig.~\ref{fig_r2r_ser_vs_bw}. A ghost target with a high intensity is observed at half-speed and half distance, i.e. $v=15$ m/s and $d=50$ m. 

If the interference comes from a LOS transmission, noting that the desired radar signal is always a backscattered signal, the signal to interference ratio (SIR) is 
\begin{align}
    \text{SIR}_{\text{R2R}} &= \frac{\gammarad^2}{\Ptxrad \gaintx \gainrx  \lambda_r^2/(4\pi d_I)^2 } = \frac{\sigma  d^2_I}{4\pi d^4}
\end{align}
where $d$ is the target distance and $d_I$ is the distance to the interferer. Since $d_I$ and $d$ are of the same order, the interference is much stronger than the desired signal. %\textcolor{blue}{If there is bistatic interference emerging from multipaths, SIR of R2R interference depends on the ratio of distances and radar cross sections, and is not as high as direct R2R interference.}

\subsubsection{Probability of R2R Interference}\label{sec_r2r_prob}

Let us assume that starting times of FMCW chirps are uniformly distributed for all vehicles. Vehicle 1 in Fig.~\ref{fig_scenarioIntR2R} transmits the radar signal $s_{\text{r}}(t)$ in \eqref{eq:TXwaveform} and Vehicle 2, which is $d$ apart from Vehicle 1, transmits its FMCW signal with a delay $\tau$ with respect to Vehicle 1. The probability of R2R interference ($P_{\text{R2R}}^{\text{int}}$) is the probability that the received signal at the victim radar lies in the band $\fr+{B_\text{r}}{T}/t+[-B_{\text{max}}, 0]$ given that the victim radar starts its transmission at time $t$. Hence, R2R interference occurs when at least one chirp of the victim radar is affected and $P_{\text{R2R}}^{\text{int}}$ is the same as the fraction of the vulnerable time over the frame time, which is explained below.

\begin{defn}[Vulnerable period $V$ \cite{CananRadConf}]
Given a victim vehicle radar that starts an FMCW transmission at time $t=0$ and a facing vehicle radar with overlapping field-of-view that starts a transmission at time $t=\tau$, the vulnerable period $V$ is the set of $\tau$ values within a chirp duration, for which interference to the victim vehicle radar occurs. 
\end{defn}
To quantify the interference, we can thus determine the vulnerable period and then compute the probability of interference occurring within the vulnerable period. 
\begin{prop}
Considering an interferer (either direct or reflected) at any distance up to $\alpha_{d}2d_{\text{max}}$ and at any relative velocity up to $ \pm v_{\text{max}}$, the vulnerable period for R2R interference is given by 
\begin{align}
   V \approx \left [-\alpha_{d} T_\text{max}, T_\text{max} \right] \label{eq:VulnerableDurationTHM}
\end{align}
where $T_\text{max} = T B_\text{max}/B_\text{r}= 2 d_\text{max}/c$, is the maximum delay of (intended) radar reflections and $\alpha_d$ is a constant determined by the longest interference path.
\end{prop}
\begin{IEEEproof}
See Appendix A.
\end{IEEEproof}
We note that the vulnerable period depends on the distance of the longest interference path. This implies that in sparse VANETS (where $\alpha_d \gg 1$), we have a long vulnerable period, but few interferers, while in dense VANETS (where $\alpha_d\le 1$ due to signal blockage), we have a short vulnerable period, but many potential interferers.  

An FMCW radar transmits $N$ successive chirps and R2R \emph{interference occurs if any two chirps of two different vehicles overlap in the vulnerable period of at least a single chirp}. Hence, any radar chirp sequence starting $(N-1)T$ prior up to the end of the radar transmission may result in R2R interference due to one or more chirps overlapping. The vulnerable period taking a whole radar frame into account is
\begin{align}
V^{(f)} = &\cup_{k=-(N-1)}^{N-1} \left[kT-\alpha_d T_\text{max},kT+T_\text{max}\right],\label{eq:Vulnerable2}
\end{align} 
 and the vulnerable duration is $\lvert V^{(f)} \rvert =  (2N-1)(1+\alpha_d)T_\text{max} \approx 2(1+\alpha_d)NT_\text{max}$, since generally $N \gg 1$.  %Hence,  any two facing radar transmissions starting radar chirp sequences at $t=0$ and $t'=\tau \in V^{(f)}$ are exposed to R2R interference. 

The probability of R2R interference among two vehicles is easily found as 
\begin{align}
\label{eq:chirpInt}
P_{\text{R2R}}^{\text{int}}=\frac{\lvert V^{(f)} \rvert}{T_f}
%= \frac{4UBmax}{Br}
=\frac{(1+\alpha_d)(2N-1) U B_\text{max}}{ NB_\text{r}} \approx\frac{2 (1+\alpha_d)U B_\text{max}}{ B_\text{r}} %-   \frac{CT B_\text{max}}{T_f  B_\text{r}}
%= CU\frac{B_\text{max}}{B_\text{r}}
%= \frac{2N C T B_max}{Br T_f}
%= \frac{2 C  B_max U }{Br}
\end{align}
 where $U=NT/T_f \in (0,1]$ is the radar duty cycle,
 %$K=T_f/(T(N+1))$,
%where $U=NT/T_f \in (0,1]$ is the radar duty cycle. 
%\frac{2NCT_\text{max}}{T_f}
%\CA{I think we should use radar duty cycle, $K_{dc}=NT/T_f$, in the rest of article also... The number of rTDMA slots per TimeSlot is defined as $K$ later on, which is $(T_f/((N+1)T))\approx1/K_{dc}$. But I did not like the notation $K_{dc}$, what can it be?}
indicating that R2R interference is minimized with reduced radar bandwidth of interest $B_\text{max}$ (or longer chirps).

\subsubsection{Example}
\begin{figure}[hb!]
\centering
\includegraphics[width=0.9\columnwidth]{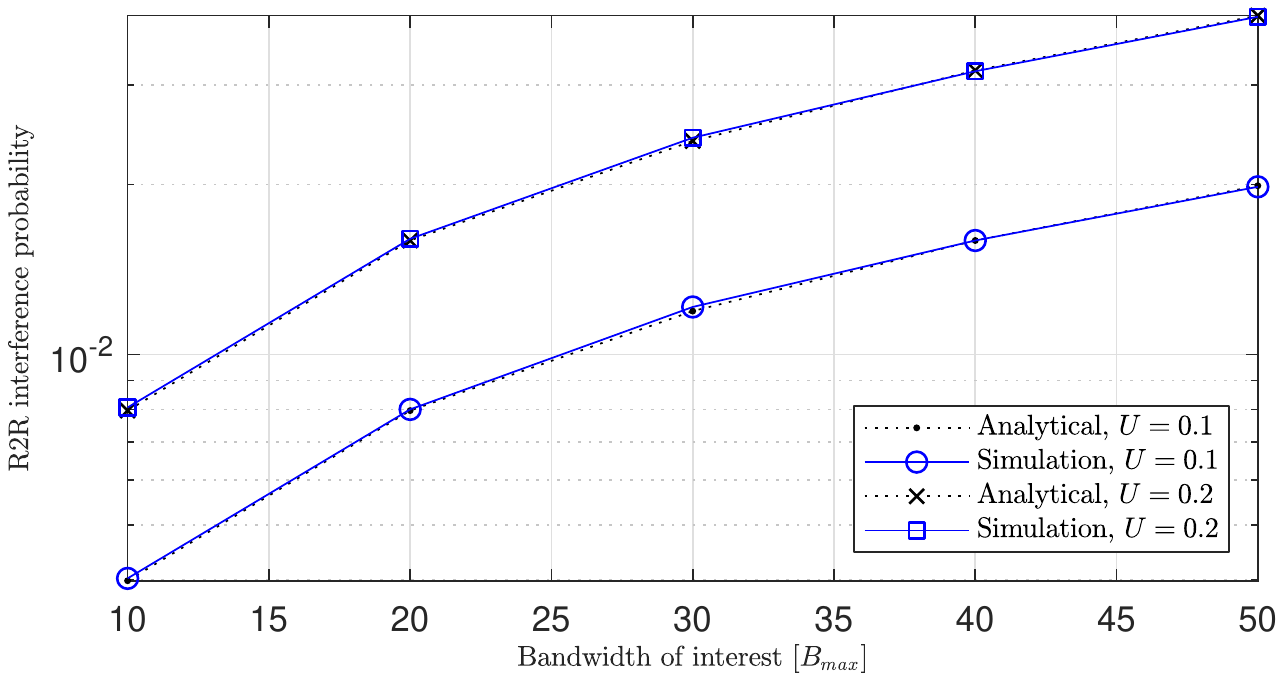}
\caption{Verification of R2R interference probability with simulations for $B_r=\SI{1}{\giga\hertz}$ and $N=99$ for two radars.}
\label{fig_R2RintProbVerification}
% \vspace{-0.2in}
\end{figure}

The R2R interference probability in (\ref{eq:chirpInt}) is verified in Fig.~\ref{fig_R2RintProbVerification} with simulations for two radars within $2 {d_\text{max}}$ distance (i.e., $\alpha_d=1$)) for varying $B_\text{max}$ and $U$. $10^6$ Monte Carlo simulations are performed using the parameters in Table \ref{tab:parameters} by assuming uniform distribution of radar starting times. For each simulation, we check if the interference is present within the bandwidth of the radar for at least one chirp within the frame. The number of occurrences of interference over the total number of simulations is the simulated interference probability. The simulations are observed to exactly match analysis in (\ref{eq:chirpInt}). A verification of the vulnerable period for various victim-interferer distances is also presented in~\cite{CananRadConf}.  

\vspace{-0.2in}

\subsection{C2R Interference Analysis}
\label{subsec:C2R}

%Interfering communication complicates detection of radar targets by increasing the noise floor over the entire range-Doppler map. In this subsection, we derive the FMCW radar received signal model in the presence of communications interference and illustrate the effect of interference on radar processing steps and performance metrics (e.g., probability of detection and probability of false alarm). 

To provide a theoretical analysis of C2R interference, we focus on a vehicular communication and sensing environment as depicted in Fig.~\ref{fig_scenarioInt}, where Vehicle~$1$ receives simultaneously the communications interference and the desired radar return from Vehicle~$2$.

\subsubsection{Impact and Power of C2R Interference}\label{sec_imp_c2r}
The interference will be spread over the entire bandwidth and lead to an increase of the effective noise floor. 
As the desired radar signal is always a backscattered signal while the interference can emanate from a direct transmission, the signal to interference ratio is
\begin{align}
    \text{SIR}_{\text{C2R}} & = \frac{\Ptxrad \gaintx \gainrx \sigma \lambda^2/((4\pi)^3 d^4)}{\Ptxcomm \gaintx \gainrx \lambdacom^2/(4\pi d_I)^2}\approx \frac{ \Ptxrad \sigma d^2_I}{\Ptxcomm 4\pi d^4}
    \label{eq_SIR_C2R}
\end{align}
When $d_I$ and $d$ are of the same order (which is typically the case if the communication transmitter of the vehicle to be detected interferes with the victim radar), the SIR expression in \eqref{eq_SIR_C2R} can be rewritten as
\begin{equation}\label{eq_sir_c2r_new}
    \sircr \approx \frac{ \Ptxrad \sigma }{\Ptxcomm 4\pi d^2}~.
\end{equation}
This means that when communication and radar use similar transmission power, the interference will be much stronger than the desired signal since $4\pi d^2 \gg \sigma$ for typical RCS values at $77 \, \rm{GHz}$ \cite{refOfMatlab}. On the other hand, the effect of C2R interference on radar performance becomes less severe when \textit{(i)} communication operates at a much smaller power than radar (i.e., $\Ptxcomm \ll \Ptxrad \sigma / (4\pi d^2) \ll \Ptxrad$), and/or \textit{(ii)} the interferer gets closer to the victim radar (i.e., for small $d$). 

\subsubsection{C2R Interference Time Ratio} \label{sec_c2r_prop_int}
We investigate the percentage of time an FMCW radar receiver is disrupted by interference from a spectrally coexistent communications transmitter (i.e., C2R interference time ratio). Let $x(t)$ denote the transmitted baseband communications signal from Vehicle~$2$, as defined in \eqref{eq_comm_signal_baseband}. Assume that Vehicle~$2$ continuously transmits data. Then, the received bandpass signal at Vehicle~$1$ radar in the presence of communications interference from Vehicle~$2$ can be written as \cite{Bliss_Inner_Bounds_TSP,joint_comm_rad_TSP_2018}
\begin{equation}\label{eq_c2r_return}
    r(t) = \rrt + \gammacom  x(t)e^{j2 \pi \left[  (\fcomm + \fDc)t - \fcomm d_0/c \right] }
\end{equation}
where $\rrt$ is the received FMCW waveform of the form \eqref{eq_radar_received}. After mixing as defined in Sec.~\ref{sec_radar_rec}, the baseband communications signal during chirp $k$ at the radar receiver is given by 
\begin{equation} \label{eq_com_baseband}
   \xctkt =  
    \gammacom x(kT + t) e^{j \theta_k} e^{j2 \pi (\fcomm + \fDc - \fr + k B_\text{r} - \frac{B_\text{r} t}{2T})t} 
\end{equation}
where $\theta_k$ is a phase that is irrelevant to the subsequent analysis. The instantaneous frequency of the baseband communications interference in \eqref{eq_com_baseband} during chirp $k$ is thus 
\begin{align}
f(kT,t)=\fcomm+\fDc-\fr + k B_\text{r} - B_\text{r} t /T  
\end{align}
with the bandwidth $\Bcomm$ as defined in \eqref{eq_comm_signal_baseband}. Hence, the radar signal is subjected to communication interference when $f(kT,t) - \Bcomm/2 \leq 0$ and $f(kT,t) + \Bcomm/2 \geq  -B_\text{max}$. In turn, this implies non-zero interference for chirp $k$ when 
\begin{equation}\label{eq_vulnerable_com}
    t \in \Tksetcr \triangleq \big\{ t ~ \rvert ~ \ttildekcr  \leq t \leq \ttildekcr + \deltatcr  \big\}
\end{equation}
over the interval $\left[ kT, \, (k+1)T \right]$, where
\begin{align}\label{eq_vulnerable_com1}
     \ttildekcr &=  \bigg( k + \frac{\fcomm + \fDc - \fr - \Bcomm/2 }{B_\text{r}} \bigg) T  \\ \label{eq_vulnerable_com22}
     \deltatcr &= \bigg(  \frac{ \min\{ B_\text{max} + \Bcomm, B_\text{r} \}  }{B_\text{r} } \bigg) T.
\end{align}
Hence, \eqref{eq_vulnerable_com} defines a \textit{communications vulnerable period} of $ \deltatcr$ seconds over a chirp duration of $T$ seconds. Different from Definition~1 in Sec.~\ref{subsec:R2R}, the radar receiver periodically suffers from this interference irrespective of the delay between radar and communication transmission times\footnote{This is the reason why C2R (and, R2C) interference effect can be characterized through \textit{time percentage} instead of \textit{probability} as in the case of R2R interference.}. Using \eqref{eq_vulnerable_com22}, the C2R interference time ratio is given by
\begin{align}\label{eq_c2r_prop_int_time}
    P_{\text{C2R}}^{\text{int}} = \frac{ \min\{ B_\text{max} + \Bcomm, B_\text{r} \}  }{B_\text{r}}U.
\end{align}
The time percentage of C2R interference can be minimized by choosing a small communication bandwidth $\Bcomm$ or small radar bandwidth of interest $B_\text{max}$ or a high radar (sweep) bandwidth $B_\text{r}$.  %The impact of the interference depends on the relative powers $\gammacom^2 / \gammarad^2$, which scales as $d^2$ as noted from \eqref{eq_radar_received_power} and \eqref{eq_gammacom}. 

\subsubsection{Example} \label{sec_ex_c2r}
\begin{figure}%[t]
\vspace{-0.2in}
\centering
\includegraphics[width=1\columnwidth]{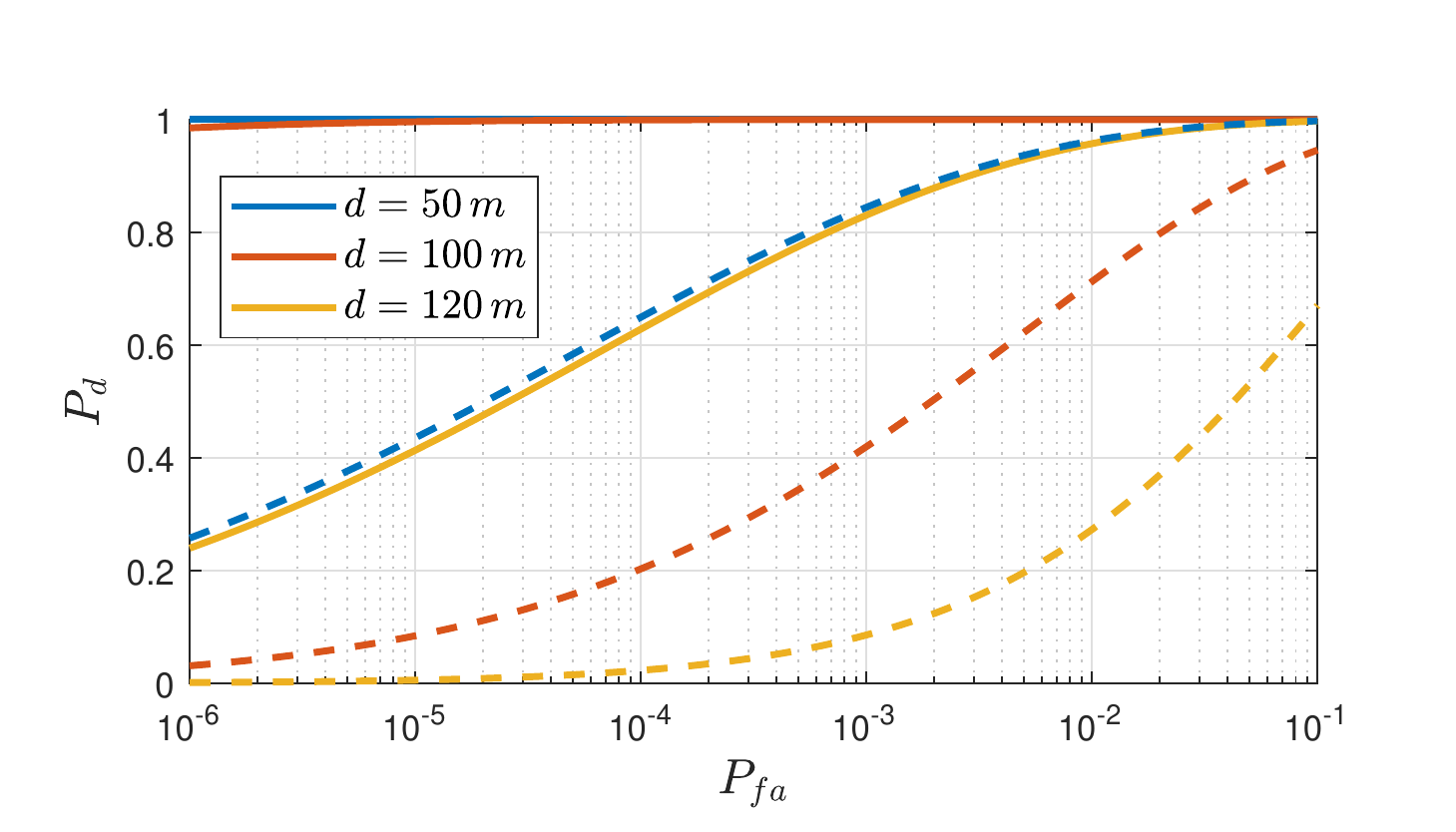}
\caption{Radar receiver operating characteristic curves for various values of distance $d$ in the absence and presence of communications interference with $d_I = d$, where $\Ptxrad = 5\,$mW, $\Ptxcomm = 5\,$mW, $\fr = 77\,$GHz, $\fcomm = 77.5\,$GHz, $\Brad = 1\,$GHz and $\Bcomm = 40\,$MHz. Solid (dashed) lines correspond to interference-free (interference) cases.}
\label{fig_c2r_pd_pfa}
\vspace{-0.2in}
\end{figure}

We demonstrate the effect of interferer distance on probability of detection, $P_\text{d}$, and probability of false alarm, $P_\text{fa}$, in Fig.~\ref{fig_c2r_pd_pfa}, where $P_\text{d}$ is calculated from \cite[Eq.~(6.36)]{richards2005fundamentals}
\begin{equation}
    P_\text{d} = \frac{1}{2} \erfc{ \left(  \erfc^{-1}\left( 2 P_\text{fa} \right) - \sqrt{\sinr} \right) }
\end{equation}
with $\sinr$ representing the signal-to-interference-plus-noise-ratio of the range-Doppler cell containing the desired target echo for a given power and distance of interference. Due to $d^2$ and $d^4$ scaling laws, respectively, for communication and radar power attenuation, an interfering car (i.e., Vehicle~2 in Fig.~\ref{fig_scenarioInt}) at a larger distance induces more severe degradation in radar detection performance. Hence, in agreement with \eqref{eq_sir_c2r_new}, spectral coexistence of FMCW radar and communication systems without significant performance reduction in radar receiver is possible only for close interferers (e.g., less than $50\,$m) in a scenario with $\Ptxrad = \Ptxcomm = 5\,$mW, $\Brad = 1\,$GHz and $\Bcomm = 40\,$MHz.\footnote{We note that the target and the interferer are co-located in this example ($d_I = d$), corresponding to the worst-case scenario in terms of radar detection. For interferers located further than the target of interest ($d_I > d$), radar performance will be less affected by C2R interference, as implied by \eqref{eq_SIR_C2R}.} 

%where 
%\begin{align}
 %   \tildernk &= \gammacom \tildex^{*}\left(kT + n T_s - \tauc - \frac{d}{c} \right)
%\end{align}

%\begin{figure}
%    \vspace{-0.5cm}
%	\begin{center}		
%		\subfigure[]{
%			\label{fig_time_freq_after_dechirping}
%			\includegraphics[scale=0.6]{figures/d_50m_P_1_mW_onesided_BW_20e6.eps}
%		}
%		\vspace{-0.4cm}
%		\subfigure[]{
%			\label{fig_time_freq_after_ADC}
%			\includegraphics[scale=0.6]{figures/d_50m_P_1_mW_onesided_BW_20e6_after_ADC.eps}
%		}
%	\end{center}
%	\caption{Time-frequency plot of a single FMCW chirp \subref{fig_time_freq_after_dechirping} after dechirping and \subref{fig_time_freq_after_ADC} after ADC low-pass filtering/sampling in the presence of communications interference, where $d = 50\,$m, $v=30\,$m/s, $\Ptxcomm = 1\,$mW, $\fcomm = 77.5\,$GHz and $\Bcomm = 40\,$MHz. The ADC low-pass filter is indicated by dashed lines in \subref{fig_time_freq_after_dechirping}.}
%	\label{fig_time_freq_plots}
%\vspace{-0.5cm}
%\vspace{-0.85cm}
%\end{figure}

%\begin{figure}%[t]
%\centering
%\includegraphics[width=1\columnwidth]{figures/d_50m_P_1_mW_onesided_BW_20e6_RangeDoppler.eps}
%\caption{Radar range-Doppler map in the presence of communications interference, where $d = 50\,$m, %$v=30\,$m/s, $\Ptxcomm = 1\,$mW, $\fcomm = 77.5\,$GHz and $\Bcomm = 40\,$MHz.}
%\label{fig_c2r_range_Doppler}
%\end{figure}

\subsection{R2C Interference Analysis}
\label{subsec:R2C}
%\todo{Symbol error rate calculations?}

In this section, we investigate R2C interference effects on communication receivers. First, we provide a received signal model in the presence of FMCW radar interference. Then, we analyze the symbol error probability of a 16-QAM system under different parameter settings.

\subsubsection{Impact and Power of R2C Interference}

The FMCW radar signal will temporarily interfere with the communication signal. The SIR is now
\begin{align}
    \text{SIR}_{\text{R2C}} & =\frac{\Ptxcomm \gaintx \gainrx \lambdacom^2/(4\pi d)^2}{\Ptxrad \gaintx \gainrx \lambda^2/(4\pi d_I)^2}\approx  \frac{\Ptxcomm d^2_I}{\Ptxrad d^2}.
\end{align}
Since $d_I$ and $d$ are of the same order and $\Ptxrad$ is generally larger than $\Ptxcomm$, the interference will be strong and cause loss of a fraction of the data.

\subsubsection{R2C Interference Time Ratio}\label{sec_r2c_prob}

Consider the R2C interference scenario in Fig.~\ref{fig_scenarioInt}, where Vehicle~2 receives communications signal from Vehicle~3 while simultaneously being corrupted by radar interference from Vehicle~1. The relative velocity of Vehicle 1 and Vehicle 2 is denoted by $v_I$ so that $d_I = d_{I,0} - v_I t$ for an initial distance $d_{I,0}$. Using a similar type of analysis as for the C2R interference, the baseband radar interference signal due to the $k$th chirp of Vehicle~1 at the communications receiver of Vehicle~2 can be written as 
\begin{equation}\label{eq_r2c_baseband}
    x_r(t, kT) = \widetilde{\gamma}_r e^{j \widetilde{\theta}_k} e^{j 2 \pi \left[ ( \fr + f_{D,I} - \fcomm )t + \frac{B_\text{r}}{2T} (t - kT - \frac{d_{I,0}}{c})^2  \right] }
\end{equation}
where $\widetilde{\gamma}_r$ and $ \widetilde{\theta}_k$ are quantities irrelevant to our analysis, and $f_{D,I} = v_I \fr /c$. The signal in \eqref{eq_r2c_baseband} is filtered out at the communications receiver when its instantaneous frequency is outside the interval $\left[ -\Bcomm/2, \Bcomm/2 \right]$. Therefore, the $k$-th chirp of the radar interferes with the communication signal during a time, $\Tksetrc$, the \textit{radar vulnerable period} at the communications receiver, which can be expressed as
\begin{equation}\label{eq_r2c_vulnerable}
    \Tksetrc \triangleq \big\{ t ~ \rvert ~ \ttildekrc  \leq t \leq \ttildekrc + \deltatrc  \big\} 
\end{equation}
where $\ttildekrc$ and $\deltatrc$ are defined as
\begin{align}\label{eq_vulnerable_com3}
     \ttildekrc &=  \bigg( k + \frac{\fcomm - \fr - f_{D,I} - \Bcomm/2 }{B_\text{r} } \bigg) T ~, \\ \label{eq_vulnerable_com_deltat}
     \deltatrc &= \frac{\min\{\Bcomm, B_\text{r} \} }{B_\text{r}} T ~.
\end{align}
The R2C interference time ratio is then given by 
\begin{align}\label{eq_prob_R2C_int}
    P_{\text{R2C}}^{\text{int}} = \frac{\min\{\Bcomm, B_\text{r} \}}{B_\text{r}} U
\end{align}
which can be minimized by choosing small communication bandwidths or large radar bandwidths.  

\subsubsection{Example}\label{sec_r2c_example}

\begin{figure}
\vspace{-0.2in}
\centering
\includegraphics[width=1\columnwidth]{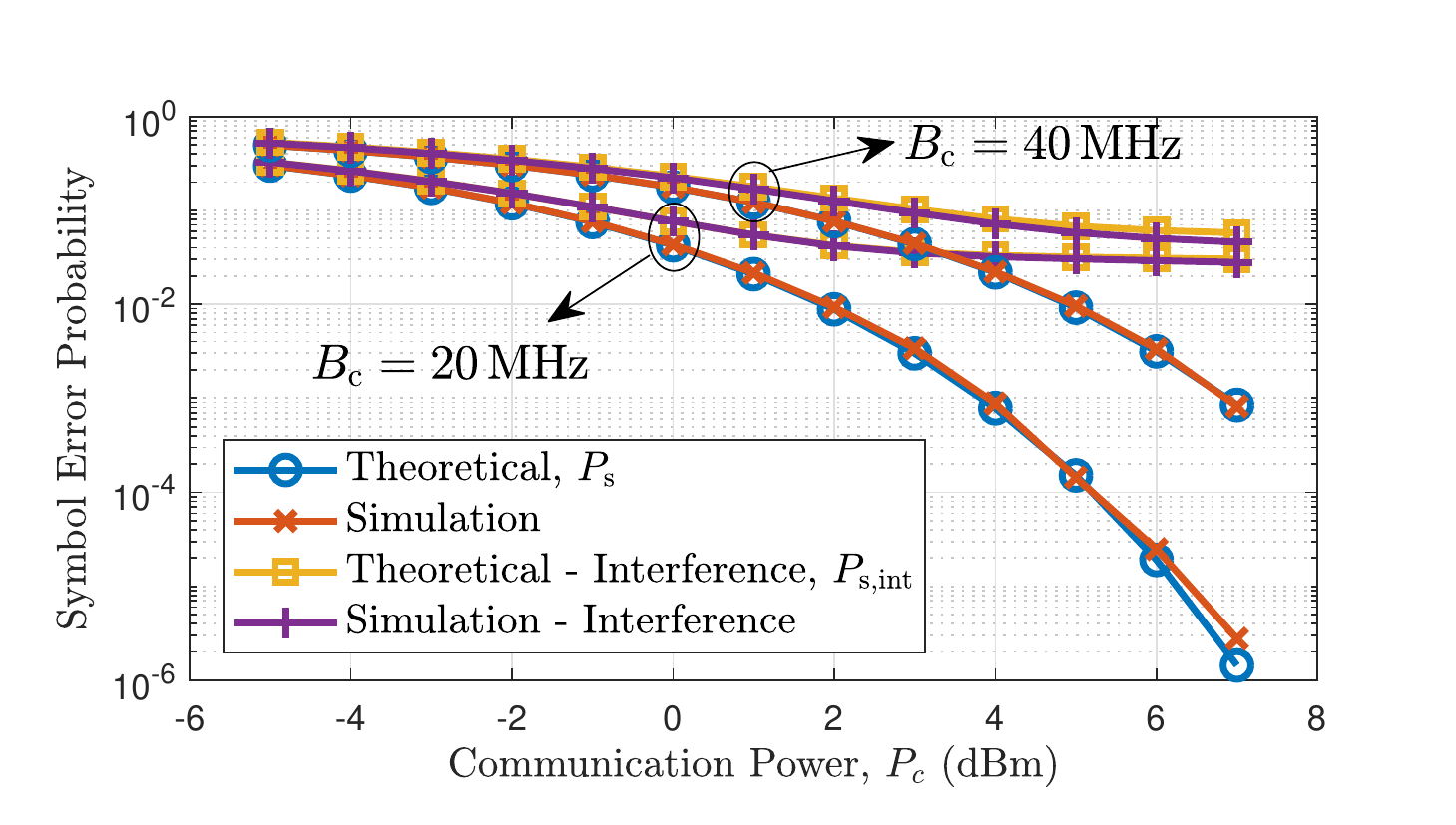}
\caption{Symbol error probability with respect to communication power for a 16-QAM scenario with and without radar interference for different values of communication bandwidth, where $d_I = d = 100\,$m, $\Ptxrad = 5\,$mW, $\fr = 77\,$GHz, $\fcomm = 77.8\,$ GHz and $\Brad = 1\,$GHz.}
\label{fig_r2c_ser_vs_bw}
% \vspace{-0.2in}
\end{figure}

\begin{figure}
\vspace{-0.2in}
\centering
\includegraphics[width=1\columnwidth]{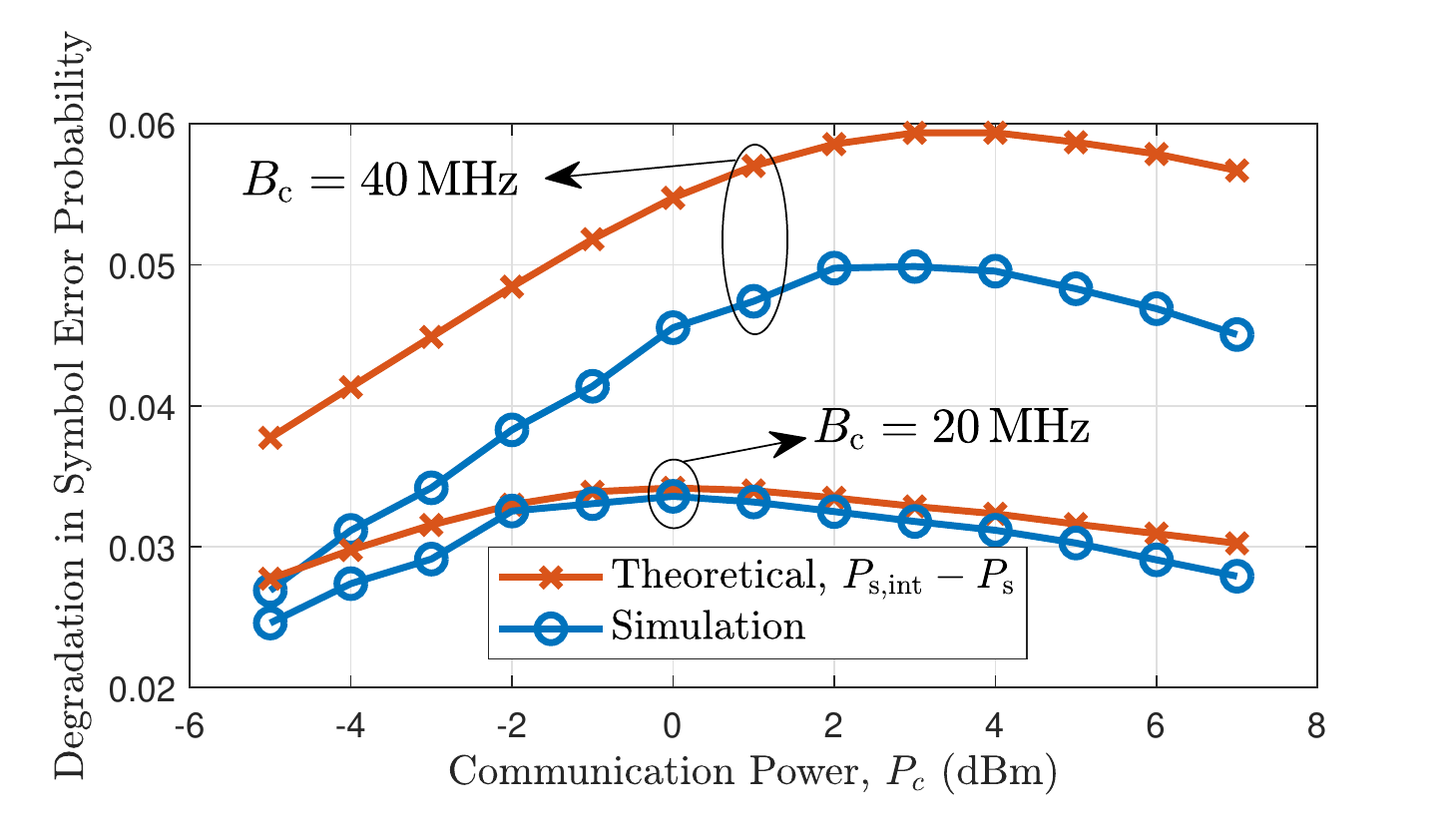}
\caption{Interference-related degradation in symbol error probability with respect to communication power for a 16-QAM scenario for different values of communication bandwidth, where $d_I = d = 100\,$m, $\Ptxrad = 5\,$mW, $\fr = 77\,$GHz, $\fcomm = 77.8\,$ GHz and $\Brad = 1\,$GHz.}
\label{fig_r2c_ser_diff_vs_bw}
\vspace{-0.1in}
\end{figure}

Based on the scenario in Fig.~\ref{fig_scenarioInt}, %Fig.~\ref{fig_r2c_plot_example} illustrates the 16-QAM received baseband communications signal in the presence of FMCW radar interference. We note that the vulnerable period characterization in \eqref{eq_r2c_vulnerable} complies with Fig.~\ref{fig_r2c_plot_example}.
we investigate the symbol error rate (SER) of the communications receiver at Vehicle 2 versus the transmit power of Vehicle 3 in Fig.~\ref{fig_r2c_ser_vs_bw} for two different communication bandwidths\footnote{In the SER performance analysis, we do not take into account any type of error correction coding. If such a coding scheme is utilized, the SER values in Fig.~\ref{fig_r2c_ser_vs_bw} can be obtained with lower communication powers, depending on coding gains achieved at different SNRs \cite[Ch.~8.1]{goldsmith2005wireless}.}. The figure shows Monte Carlo simulation results of a 16-QAM system with and without radar interference along with the (semi-)analytically derived $\Pswoint$ and $\Psint$ values, where $\Psint$ constitutes an upper bound on the theoretical SER value in the presence of radar interference, computed as\footnote{This is similar to compound rate in \cite{joint_comm_rad_TSP_2018}.}
\begin{equation}\label{eq_Ps_int}
    \Psint = \alphaint + \Pswoint \, \left(1 - \alphaint \right)
\end{equation}
with $\Pswoint$ and $\alphaint$ denoting, respectively, the SER of 16-QAM without radar interference \cite[Eq.~(6.23)]{goldsmith2005wireless} and the ratio of symbols that are subjected to interference\footnote{\label{fn_alpha_int}More specifically, $\alphaint$ is defined as the percentage of time the magnitude of the received FMCW waveform is large enough to cause symbol error. In the low signal-to-interference-plus-noise ratio (SINR) regime, $\alphaint$ can be approximated as $\alphaint \approx \Bcomm/B_\text{r}$ (as suggested by \eqref{eq_prob_R2C_int}), while it converges to zero asymptotically at high SINR.}. We also show $\Psint - \Pswoint$, i.e., SER performance degradation due to interference, in Fig.~\ref{fig_r2c_ser_diff_vs_bw}. From the figures, it is observed that larger communication bandwidths lead to higher interference-related degradation in SER performance due to an increase in the R2C interference time ratio, as stated in \eqref{eq_prob_R2C_int}. In addition, the effect of radar interference becomes more severe as the SINR increases, in the sense that additional communication power in the interference case that is required to attain the same SER value as the interference-free case gets larger. At very high SINR values, $\alphaint$ would become zero, implying that the SER performance of the interference case converges to that of the interference-free case. We conclude from Fig.~\ref{fig_r2c_ser_vs_bw} that in a scenario where radar and communication transmitters operate at the same power (i.e., $\Ptxrad = \Ptxcomm = 5\,$mW), R2C interference from an FMCW radar with $\Brad = 1\,$GHz  leads to unacceptably high SER levels at a communication receiver with $\Bcomm = 20\,$MHz and $\Bcomm = 40\,$MHz. 

%\textcolor{red}{(e.g., $\Pswoint \approx 10^{-6}$ and $\Psint \approx 10^{-2}$ for $\Bcomm = 20\,$MHz)}.

%radar interference leads to a negligible performance degradation in the low SINR regime for all bandwidths (satisfying $\Bcomm \ll \Brad$) since $\alphaint$ is very small compared to $\Pswoint$ (see \eqref{eq_Ps_int})

%\begin{figure}
   % \vspace{-0.5cm}
%	\begin{center}		
%		\subfigure[]{
		%	\label{fig_r2c_plot_example_amp}
		%	\includegraphics[scale=0.6]{figures/fcom_77p8e9_Prad_500_mW_onesided_BW_20_MHz_time_amp.eps}
%		}
%		\vspace{-0.4cm}
%		\subfigure[]{
%			\label{fig_r2c_plot_example_amp_freq}
%			\includegraphics[scale=0.6]{figures/fcom_77p8e9_Prad_500_mW_onesided_BW_20_MHz.eps}
%		}
%	\end{center}
%	\caption{
	%\subref{fig_r2c_plot_example_amp} 
%	Time-amplitude 
	%and \subref{fig_r2c_plot_example_amp_freq} time-frequency 
%	plot for communications signal and radar interference at the receiver of Vehicle~2 for a single FMCW chirp duration, where $\fcomm = 77.8\,$ GHz, $\Bcomm = 40\,$ MHz, $\Ptxrad = 500\,$mW, $\Ptxcomm = 5\,$mW and $d_{12} = d_{23} = 140\,$m (with $d_{ij}$ representing the distance between Vehicle~$i$ and Vehicle~$j$).}
%	\label{fig_r2c_plot_example}
%\vspace{-0.5cm}
%\vspace{-0.85cm}
%\end{figure}

%\begin{figure}
%\centering
%\includegraphics[width=1\columnwidth]{figures/SER_Comm_Power_Theoretical.eps}
%\caption{Convergence of theoretical symbol error probabilities in interference-free and interference cases at high SINR.\HW{Not sure if we should keep this figure}}
%\label{fig_r2c_ser_vs_pow_theory}
%\end{figure}

\subsection{Interference Analysis Outcome}
The communication SER performance loss due to radar interference (R2C) is non-negligible for $\Ptxcomm/\Ptxrad \approx 1$ with $B_c=20$ MHz and $B_c=40$ MHz (Fig.~\ref{fig_r2c_ser_vs_bw}). 
%The SER levels achieved for communication under radar interference (R2C) are high for any $\Ptxcomm/\Ptxrad$ ratio (Fig.~\ref{fig_r2c_ser_vs_bw}), especially for $B_c=40$ MHz. 
The SER performance can be enhanced by increasing the $\Ptxcomm/\Ptxrad$ ratio, which, in turn, leads to degradation of $\text{SIR}_\text{C2R}$ (Table~\ref{tab:InterferenceTable}) and thus radar detection performance (Fig.~\ref{fig_c2r_pd_pfa}). 
%A good symbol error rate for communication could be achieved by lowering the $\Ptxcomm/\Ptxrad$ ratio, which will this time increase the $\text{SIR}_\text{C2R}$ (Table~\ref{tab:InterferenceTable}). 
This suggests that FMCW radar sensing and communication with similar powers should not occupy the same time-frequency resources. Additionally, R2R interference can be avoided if FMCW chirp sequences start transmission during non-overlapping vulnerable periods.

\section{RadChat: Protocol Description}
\label{sec_model}

\begin{figure}
\centering{}\includegraphics[width=0.7\linewidth]{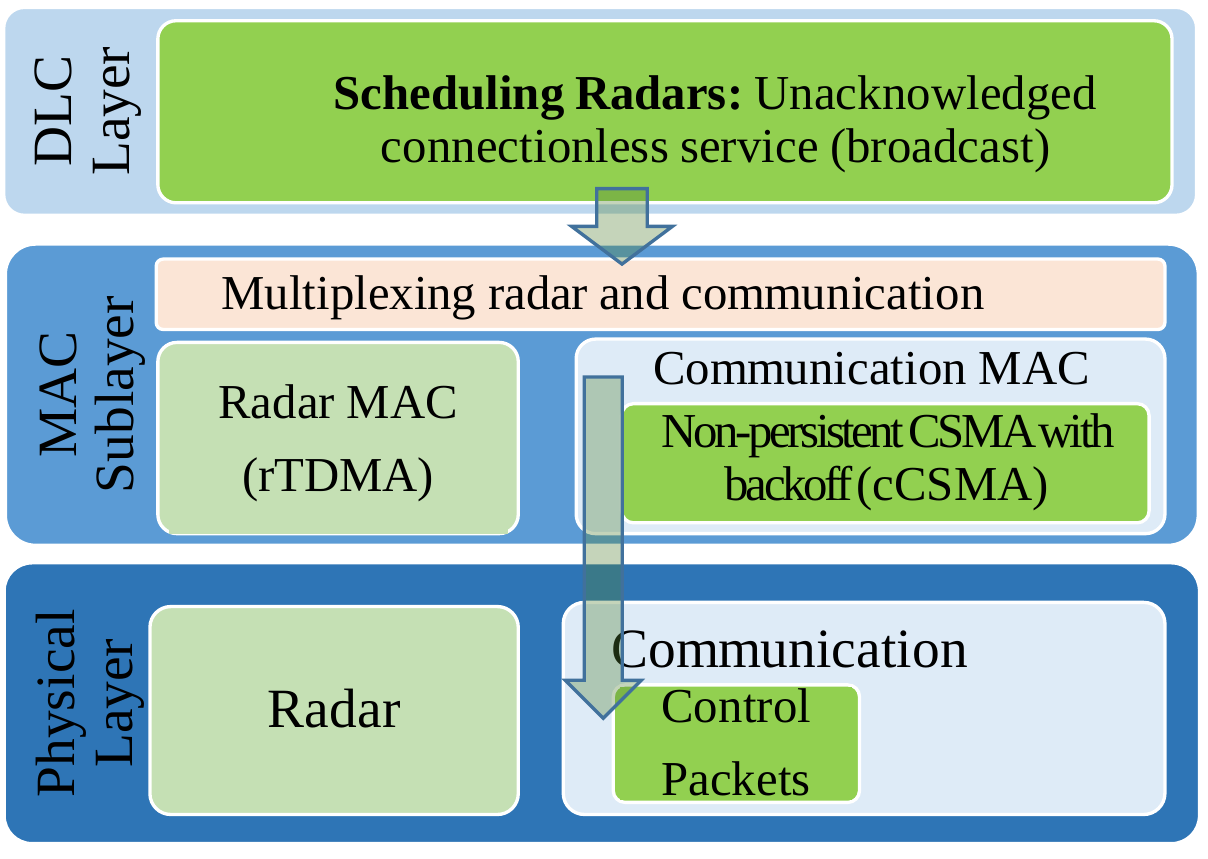}
\caption{Summary of the layered architecture of RadChat.}
\label{fig_GreenLocLayers}
\end{figure}

RadChat is proposed as a distributed radar and communication cooperation protocol, which avoids R2R interference by scheduling radar sensing to non-overlapping vulnerable periods and avoids R2C/C2R interferences by using a separate communication control channel in order to ensure
non-conflicting time-frequency blocks for communication and radar. The layered architecture of RadChat is summarized in Fig.~\ref{fig_GreenLocLayers}. 
A RadChat unit is responsible of data link control (DLC), MAC sublayer and PHY layer functions. Upper layer functions are assumed to be processed at a central unit at each vehicle, which combines data of all RadChat units, which are co-located on a vehicle looking toward different directions. 
The main service provided by the DLC layer is scheduling of radars, i.e. assigning non-overlapping vulnerable periods to radars, and is an unacknowledged connectionless service. Broadcast control packets are used to schedule radar packets\footnote{A RadChat unit is built upon a packet-switched architecture, where an FMCW radar chirp sequence is regarded as a \textit{radar packet}.} at non-overlapping time slots. 
%(rTDMA). The medium access technique used for these communication control packets is non-persistent CSMA with backoff (cCSMA). A best effort service with no acknowledgements are employed to schedule radars to different rTDMA slots. 
In addition to scheduling radars, the DLC can provide basic communication services, but these are outside the scope of the current work. 

\begin{figure}
\centering{}\includegraphics[width=1\columnwidth]{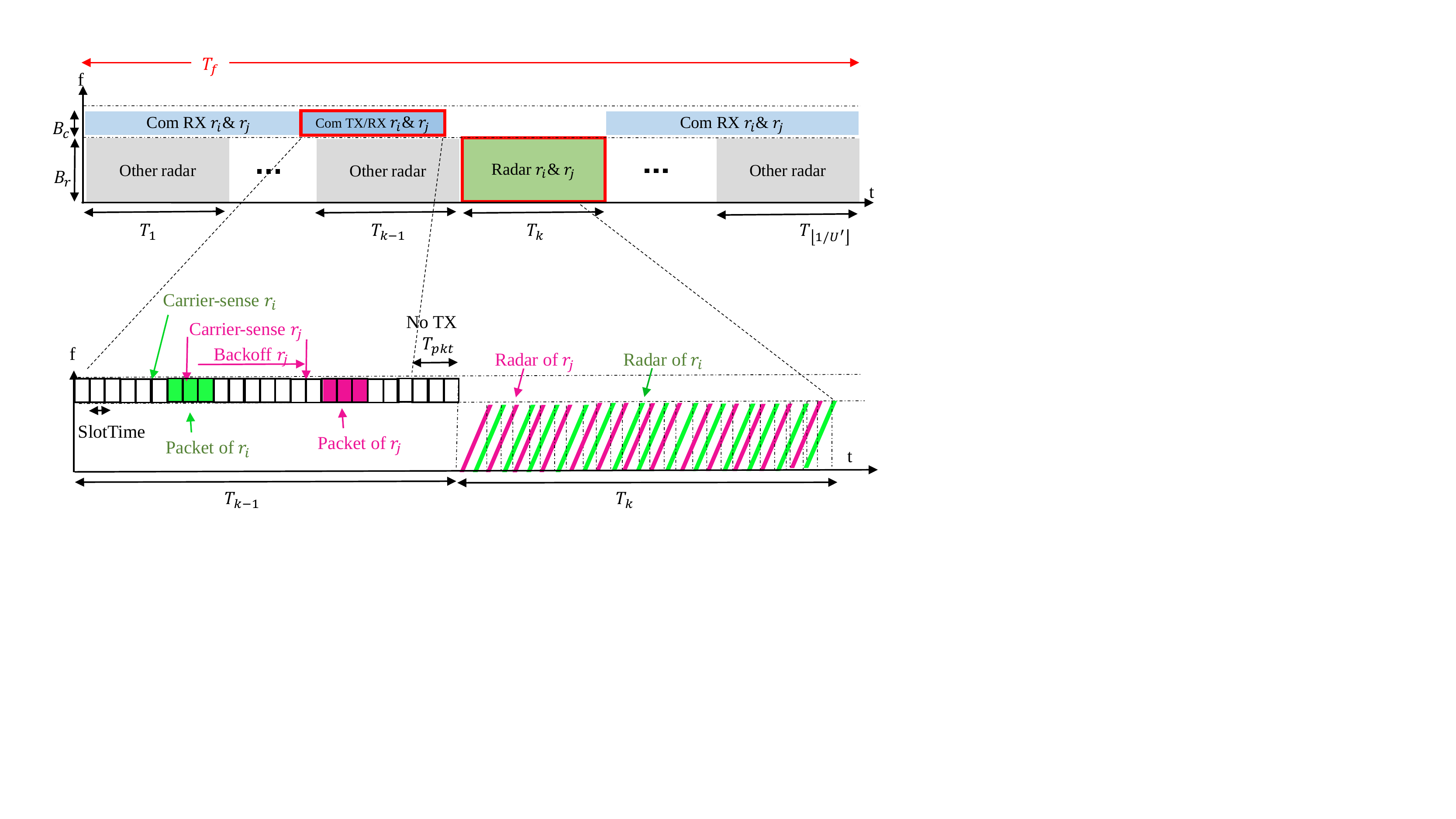}
\caption{RadChat Scheduling Radars scheme: FDM / rTDMA / cCSMA}
\label{figConf}
\end{figure}

Given that the PHY layer operates as described in the previous sections, the MAC layer of RadChat operates by FDM/rTDMA/cCSMA, which is a scheme based on frequency division multiplexing (FDM) between radar and communication, time division multiple access for radar transmissions (rTDMA), and carrier sense multiple access for communications (cCSMA). Fig.~\ref{figConf} illustrates the division of the frequency-time domain for the proposed DLC service w.r.t. two specific RadChat units $r_i$ and $r_j$. % whereas \textcolor{blue}{Fig.~\ref{figFlowchart} is a flowchart illustrating the MAC}. 
Radars are scheduled by assigning each a different rTDMA slot, where rTDMA slots are defined as radar slots with disjoint vulnerable periods. 
%\begin{figure}
%\centering{}\includegraphics[width=1\columnwidth]{figures/ITS_flowchart.pdf}
%\caption{Flowchart illustrating RadChat}
%\label{figFlowchart}
%\end{figure}

\subsection{Terminology and Assumptions}
\subsubsection{Network}
For a general but fixed VANET topology, %let $\mathcal{G}$ be a undirected connectivity graph with $\mathcal{G}=(\mathcal{V},\mathcal{E})$, 
%where % each vertex corresponds to a
RadChat unit $r_{i}$ %\in\mathcal{V}$ and each edge $e_{ij}$ means that the RadCom unit $r_i$ 
is \textit{connected to/facing} the RadChat unit $r_j$ if $r_i$ is able to receive/transmit communication signals from/to $r_j$, assuming symmetric communication links. Links may be established through LOS or reflected paths. Let $S_\text{X}$ denote the set of RadChat units mounted on vehicle $X$ %A given vehicle $X$ may be equipped with multiple RadCom units
, which are connected to one central processing unit at vehicle $X$. Each $r_i \in S_\text{X}$ uses a different rTDMA slot to handle R2R self-interference, which is the R2R interference among radars mounted on the same vehicle.
All vehicles use a common radar band $B_{r}$ and a common communication band $B_{c}<B_\text{max}$, in order to be able to reuse the radar ADC. We assume equal radar and communication transmit powers ($\Ptxcomm=\Ptxrad$) for simplicity and to ensure that RadChat provides long enough communication range to communication with all interferers within distance $\alpha_d 2 d_{\text{max}}$. 
%
%longer communication range than radar range. \textcolor{blue}{We also assume a dense VANET, where direct LOS interferers are blocked beyond $2d_\text{max}$}. Since any radar interference with a larger propagation delay than $T_\text{max}$ will have an IF frequency larger than $B_\text{max}$ and filtered out at radar receiver, $\Ptxcomm=\Ptxrad$ guarantees that any potential R2R interferer, the received power of which is high enough to be detected by the radar, is communicated. 
Extensions to dynamically changing topologies, as well as power control, which adapts these powers independently for all RadChat units, are left for future work.  
 
%
%We assume that the control packets are always sent at peak power, which is equal to radar peak power. This ensures that two radars communicate to mitigate interference as soon as they hear each other, before the radar sensor can detect the vehicle.
%

\subsubsection{Timing}
We introduce time slots $T_k$ (Fig.~\ref{figConf}) of duration $(N+1)T\le T_f$, which corresponds to the duration of $N$ chirps plus one idle chirp time accounting for the overflow of time shifted rTDMA slots. Let $U'=(N+1)T/T_f$ be defined as the modified radar duty cycle, then a radar frame is divided into $1/U'$ time slots. This slotted time is set to provide non-overlapping chirp sequences within a radar frame and thereby maximize the number of vehicles with no mutual interference, denoted by $M_\text{max}  \leq \lfloor 1/U'  \rfloor \left \lfloor {B_\text{r}}/{((1+\alpha_d) B_\text{max})} \right \rfloor $. Each time slot $T_k$ is further divided to slots called \textit{SlotTime}s of duration $\delta$, which are large enough to detect channel activity by the carrier-sensing function of the CSMA mechanism. Vehicles are assumed to synchronize their clocks using GPS.
\subsubsection{Data Structure}
Each RadChat unit $r_i$ on vehicle $X$ has several MAC state variables that are broadcast to other vehicles: 
\begin{itemize}
    \item $r_i.\text{ID}$: an identifier of the time reference, initialized to the vehicle index $X$. 
    \item $r_i.\text{SI}$: an rTDMA slot index in the local time reference, initialized to $0$. During operation, $r_i.\text{SI}\in \{ 1,\ldots, M_{\text{max}}\}$. 
    \item $r_i.t_\text{rs}$: a radar start time, initialized uniformly in $[0,T_f]$ incremented by $T_f$ every frame. 
    \item $r_i.\text{strength}$: a priority indicator, coupled to $r_i.\text{ID}$, initialized to $0$.
\end{itemize}
All RadChat units mounted on the same vehicle use the same ID and strength values, whereas SI and $t_\text{rs}$ are specific to each RadChat unit assigned by the central processor at vehicle $X$.
We will denote the set of all of the rTDMA slot indices used by RadChat units mounted on the vehicle X by $S_\text{X}.\text{SI}$, whereas the set of all of the radar start times are denoted by $S_\text{X}.t_\text{rs}$.

Due to the distributed nature of the algorithm, each vehicle $X$ will assign rTDMA slots according to its own time frame initially. The couple $(r_i.\text{ID},r_i.\text{SI})$ specifies a unique rTDMA slot index for all radars that have the same time reference $r_i.\text{ID}$. The variable $r_i.\text{strength}$ is used to give priority to the time reference which is shared the most in the network, in order to avoid fluctuations among different time references. 

Communication functions are common to all $r_i \in S_\text{X}$. Hence, a base RadChat unit $r_i^* \in S_\text{X}$ on vehicle $X$ is selected  according to which timing of communication functions are conducted. 
Each vehicle also keeps track of these MAC state variables of the base RadChat unit  $r_i^*$ during operation, which are not broadcast: 
\begin{itemize}
    \item $r_i^*.\text{counter}$: a binary exponential backoff (BEB) counter, initialized by a random integer
    $\text{rand}([0,2^{b} W_0-1])$, where $b$ is the backoff stage and $W_0$ is the maximum contention window size; $b$ is incremented upon each busy carrier sense until $b\leq B$, where $B$ is the maximum backoff stage. $b$ is reset at the end of $T_{k-1}$.  
    \item $r_i^*.t_\text{cs}$: a communication starting time, initialized to  $r_i^*.t_\text{cs}=r_i^*.t_{\text{rs}}-(N+1)T -T_\text{pkt}+\delta r_i^*.\text{counter}$, where $T_\text{pkt}$ is the duration of a control packet. $r_i^*.t_\text{cs}$ is updated whenever the radar start time of the base RadChat unit on vehicle $X$, is changed. 
\end{itemize}

\subsection{MAC Operation}
%The MAC sublayer of RadChat is composed of two distinct protocols to realize the two different kinds of DLC services:
%\begin{enumerate}
%\item TFDM/rTDMA/cCSMA for scheduling radars service. 
%\item Random rezervation based data communications service. 
%\end{enumerate}
%Radar and communication are time-frequency domain division multiplexed (FDM) with time division multiple access for radar signals (denoted rTDMA). Two different MAC schemes are used for the two distinct DLC services: (i) carrier-sense multiple access for communication signals (denoted cCSMA) is used for scheduling radar transmissions to non-overlapping rTDMA slots and (ii) a rezervation-based protocol is used for data.  
%The focus of our previous works in~\cite{CananRadConf,CananPimrc} is the first function, whereas we include the second function in this article.

%\subsubsection{TFDM/rTDMA/cCSMA for scheduling radars}
%The goal is to disseminate non-overlapping rTDMA slots among vehicles to mitigate interference. This is an FMCW-based-RadCom system, which uses time-frequency domain division multiplexing (TFDM) for radar and communication signals for a single vehicle and employs time division multiple access among radar signals of various RadCom units (denoted rTDMA). 

%rTDMA is the radar medium access scheme, where \textit{facing} radars of different vehicles are assigned non-overlapping vulnerable periods.

%RadChat uses a best-effort approach by unacknowledged communications since communication links may lose symmetry due to highly mobile VANET structure. 
In order to assign non-overlapping rTDMA slots among facing RadChat units, non-persistent CSMA with BEB is employed. rTDMA slots in $T_k$ are generally determined by communication contention during slot $T_{k-1}$ (Fig.~\ref{figConf})\footnote{Note that the contention time during slot $T_{k-1}$ is shifted to the left by $T_\text{pkt}$ in Fig.~\ref{figConf}, since any control packet transmitted at least $T_\text{pkt}$ prior to radar mode can be received.}. 
Control packets are transmitted if the channel is sensed idle for one \textit{SlotTime} $\delta$ or random BEB is employed if channel is sensed busy. %Details and the state diagram of the radar scheduling service of RadChat is given in~\cite{CananRadConf}.  
%
%\subsubsection{Initialization}
%Upon initialization, each RadCom unit $r_i \in U_X$ is assigned a radar start time ($t_{rs}$) and a random backoff counter value ($b$). A  communication control packet starting time $t_{cs}$ is calculated as given in Algorithm ~\ref{algo_init}, where $T_{pkt}$ is the control packet duration. $SI$ is the rTDMA slot index, which is initially $-1$. The rTDMA slots are numbered from 1 to $M_\text{max}$. %Within each vehicle, all RadCom units are assumed to use the same $SI$ due to directivity of RadCom antennas. 
 %Hence each rTDMA slot index is unique together with the identity of the vehicle. $ID$ is the vehicle identity, and ($ID$,$SI$) represents a unique rTDMA slot index.   
%\begin{algorithm}
%\caption{Initialization}\label{algo_init}
%\begin{algorithmic}[1]
%\Procedure{Initialize(Vehicle $X$)}{}
%\For {$r_i \in U_X$}
%\State $r_i.t_{rs} \gets \text{assign radar starting time}$
%\State $r_i.SI \gets -1$
%\State $r_i.ID \gets X$
%\State $r_i.strength \gets 0$
%\State $r_i.b \gets \text{rand}([0,2^{b_{CW}} CW_{\text{max}}-1])$
%\State $r_i.t_{cs} \gets r_i.t_{rs}- %T_{pkt}-(N+1)T+r_i.b$.
%\EndFor
%\EndProcedure
%\end{algorithmic}
%\end{algorithm}
%
Each vehicle $X$ may prefer to allocate all radar transmissions of its mounted RadChat units in the same time slot $T_k$, $S_\text{X}.t_{\text{rs}} \in T_k$ (if the number of RadChat units on a single vehicle is $\leq  \left \lfloor {B_\text{r}}/{((1+\alpha_d) B_\text{max})} \right \rfloor$). It is not necessary to squeeze all radars of a vehicle to a single time slot $T_k$, however at least one time slot $T_{k-1}$ should be empty with no radars, so that it can be used for communication jointly by all $r_i \in S_\text{X}$.

Each vehicle $X$, which has a set of random radar start times $S_\text{X}.t_{\text{rs}}$, selects a contention period, preferably the prior time slot $T_{k-1}$ where most of the radar start times reside in $T_k$ and selects a base RadChat unit $r_i^*$, $r_i^*.t_\text{rs}\in T_k$ according to which communication start time $r_i^*.t_{\text{cs}}$ is calculated. $\forall r_i \in S_\text{X}$ transmit a single communication control packet during $T_{k-1}$. This control packet is broadcast to all RadChat units connected to the Vehicle $X$ (as if omni-directional communication) and contains the following information: identity of the transmitter ($r_i$), time reference frame ($r_i.\text{ID}$) and the set of all rTDMA slot indices of RadChat units mounted on vehicle $X$  ($S_\text{X}.\text{SI}$), strength of this time reference frame ($r_i.\text{strength}$), the base RadChat unit's radar starting time and its slot index ($r_i^*.t_{\text{rs}},r_i^*.\text{SI}$). MAC functions of RadChat are presented next.
\begin{itemize}
\item \emph{RadChat Carrier Sensing:} 
{A RadChat unit $r_i$  intends to start radar transmission at $r_i.t_{\text{rs}}\in T_k$ in Fig.~\ref{figConf}. This RadChat unit carrier-senses the communication channel $\Bcomm$ during the entire radar frame except during $T_k$ (as it is transmitting radar), and receives incoming control packets.} %We assume that all RadCom units on a vehicle X carrier sense a communication either by LOS path or through reflections in the VANET. In real settings, the directivity of RadCom units may result with different carrier sensing outcomes among RadCom units within a vehicle. Then, the aggregated carrier sensed channel is taken into account in CSMA operation. In this study, we investigate the worst case situation where all communication activity is carrier sensed by all RadCom units within a vehicle.}
\item \emph{RadChat Transmission at $T_{k-1}$:} 
If a control packet transmission is scheduled at $r_i^*.t_\text{cs} \in T_{k-1}$, 
 carrier sensing is employed during  $T_{k-1}$. The control packet is sent if channel is sensed as idle, or backoff is employed if channel is sensed as busy (and $r_i^*.t_\text{cs} \gets r_i^*.t_\text{cs}+\delta r_i^*.t_\text{counter}$).
Upon completion of transmission of a control packet by the RadChat unit $r_i$, 
if $r_i.\text{SI}=0$, $r_i$ updates  $r_i.\text{SI}$ to the assigned value by the central processor. % of the vehicle  if $r_i.\text{SI}=0$, the value is set to the assigned value by the central processor of the vehicle, otherwise it is left unchanged. 
%If no control packet is transmitted during $T_{k-1}$ and time to begin radar transmission has come, i.e., $r_i^*.t_{\text{cs}}+T_{\text{pkt}} \notin T_{k-1}$, then the transmission of a control packet is skipped for the current radar frame and scheduled for the next frame time.
\item \emph{RadChat Reception at $T_{k-1}$:} Upon reception of a control packet from $r_i$ by $r_j$  (which was not transmitting radar at that time), $r_j$ updates its state as described in Algorithm ~\ref{algo_main}. Throughout the operation of RadChat, each RadChat units stores the received ID, SI and strength information in a local database $\mathcal{D}_j$. This is used to keep track of unused rTDMA slots for a time reference, and the priority of the time reference. In lines 5, 10, and 15 the SI should be selected within $T_k$ if available, otherwise from the set of unused rTDMA slots in $T_f$. This algorithm ensures that $r_j.\text{SI}$ is assigned so that $r_j.\text{SI} \neq S_\text{X}.\text{SI} \cup \mathcal{D}_j$. $r_j.t_{\text{rs}}$ and $r_j.t_{\text{cs}}$ in Line 16 are set according to, 
\begin{align}\nonumber
        r_j.t_{\text{rs}}\gets & r_i.t_{\text{rs}}+(N+1)T \{ K_j -K_i \} %\\\nonumber
        + \vert V \vert \{\kappa_j-\kappa_i) \} \nonumber\\
        r_j.t_{\text{cs}}\gets& r_j^*.t_{\text{rs}}-(N+1)T-T_\text{pkt}+\delta r_j^*.\text{counter}, \nonumber 
    \end{align}
with $\kappa_j$ $=$ $\text{mod}(r_j.\text{SI},U'M_\text{max})$, $K_j=\lceil r_j.SI/(U'M_\text{max})\rceil$, 
where $r_j^*$ is the a base RadChat unit of the receiving vehicle.  
\end{itemize}

\begin{algorithm}
\caption{Process control packet at unit $r_j$}\label{algo_main}
\begin{algorithmic}[1]
\State Store $(r_i.\text{ID},r_i.\text{SI},r_i.\text{strength})$ in $\mathcal{D}_j$ 
\If {$r_j.\text{SI}=0$}
    \State $r_j.\text{ID} \gets r_i.\text{ID}$
    \State $r_j.\text{strength} \gets r_i.\text{strength}+1$
    \State $r_j.\text{SI} \gets  \text{SI}  \in T_k \cup T_f \setminus \{S_\text{X}.\text{SI}\} $
    %\State Store $(r_i.\text{ID},r_i.\text{SI},r_i.\text{strength})$ in $\mathcal{D}_j$ 
\Else
    \If {$r_j.\text{ID}=r_i.\text{ID}$} \Comment{same time reference}
        \State $r_j.\text{strength} \gets \max(r_j.\text{strength},r_i.\text{strength})+1$
        %\State Store $(r_i.\text{ID},r_i.\text{SI},r_i.\text{strength})$ in $\mathcal{D}_j$ 
        \If{$r_j.\text{SI}\in S_\text{X}.\text{SI}$}
            \State $r_j.\text{SI} \gets  \text{SI}  \in T_k \cup T_f \setminus \{r_i.\text{SI}| i \in \mathcal{D}_j\} $
        \EndIf
    \Else \Comment{different time reference}
        %\State $\text{Store $ID$, $SI$, $strength$}$ of $r_i$
        \If{$r_i.\text{strength}>r_j.\text{strength}$}
            \State $r_j.\text{ID} \gets r_i.\text{ID}$
            \State $r_j.\text{strength} \gets r_i.\text{strength}+1$
            \State $r_j.\text{SI} \gets  \text{SI}  \in T_k \cup T_f \setminus \{S_\text{X}.\text{SI}\} $
        \EndIf
    \EndIf
\EndIf
\State Calculate $r_j.t_{\text{rs}}$ and $r_j.t_{\text{cs}}$ if $(r_j.\text{ID},r_j.\text{SI})$ has changed
\State Update state of all other RadChat units on same vehicle
\end{algorithmic}
\end{algorithm}
%\subsubsection{Reservation-based MAC for Data Communications}
%\label{sec_cRadCom}

%The primary goal of RadChat is to mitigate R2R interference and the second goal is to enable data communications. Data communications may take place if the time-frequency domain is not fully occupied after having scheduled radars to disctinct rTDMA slots, i.e:
%\begin{equation}
%    \sum_{\text{all }r_i \in U_X} \vert S_i \vert < M_{\text{max}}. 
%\end{equation}
%The communication transmit power is assumed to be controlled so as to be received by minimum radar receiver sensitivity. 

%Under these conditions, a specific portion of the radar frame within $\Bcommtrl$ is reserved for data communications contention. These can be several Time slots before unused radar slots. A vehicle willing to send a data packet selects a random unused radar slot and sends a reservation request, which includes the carrier frequency  $f_i \in B_r$ to be reserved, source and destination IDs. If this packet is successfully received, the destination sends its acknowledgement over the reserved frequency band $\Bcomm$ with carrier frequency $f_i$. Hence, reservation packets are send over $B_{ctrl}$ during $T_{k-1}$ to rezerve $f_i \in B_r$ over a bandwidth of $\Bcomm$ and acknowledgement and data packet are sent during $T_{k}$ with carrier frequency $f_i \in B_r$ over a bandwidth of $\Bcomm$.  

\subsection{Properties of RadChat}
For a fixed connected network topology and with less than $M_\text{max}$ active radars, RadChat is guaranteed to eventually converge to a solution where each vehicle uses a distinct rTDMA slot and thus R2R interference is eliminated, when the following conditions are met: 
\begin{itemize}
    \item The radar duty cycle of RadChat must satisfy $U' \leq 1/3$. Since RadChat units cannot receive control packets when radar is active, a higher radar duty cycle may end up with two disjoint interfering networks. Higher duty cycle necessitates the use of a separate communication module for mitigation of FMCW radar interference with RadChat.
    \item RadChat allows synchronization errors of at most $\vert V\vert/2$ across vehicles, since it places each radar transmission in the middle of a vulnerable period leaving a guard time. Under perfect synchronization, RadChat can allocate up to $2M_\text{max}$ non-interfering RadChat units by a time spacing of $\vert V\vert/2$ (so that non-overlapping green-indicated radar bandwidths of $B_\text{max}$ fill the whole time-frequency domain in Fig.~\ref{fig_chirpJournal}). 
    \item Bandwidth reserved for communication $\Bcomm$ should allow for at least one data packet during $T_k=(N+1)T$:
    \begin{align}
        (N+1)T &> T_\text{pkt}=\frac{8N_\text{pkt}/\log_2(|\Omega|)}{\Bcomm/(1+\alpha)}\\
        \Bcomm&>\frac{8N_\text{pkt}(1+\alpha)}{(N+1)T\log_2(|\Omega|)}
    \end{align}
    where $|\Omega|$ is the constellation size.
\end{itemize}
Some properties of RadChat protocol are as follows:
\begin{itemize}
    \item RadChat takes care of both R2R interference among vehicles and among radars mounted on the same vehicle, i.e., self-interference.  
    \item RadChat eliminates any potential R2R interference within a distance $2 \alpha_d d_{\text{max}}$, provided $\Ptxcomm=\Ptxrad$ is sufficient to ensure communication with these interferers. Interferers beyond $2 \alpha_d d_{\text{max}}$ or beyond the maximum communication range cannot be eliminated by RadChat. 
    %that the potential interferers with propagation delays \textit{less} than the maximum radar round trip time $T_\text{max}$ are communicated and allocated different rTDMA slots. The potential interferers with propagation delays \textit{more} than $T_\text{max}$ are either filtered out at radar receiver circuity (if a perfect low-pass filter with bandwidth $B_\text{max}$ is used) or have low received radar signal power. 
    \item RadChat ensures that R2R mitigation is completed in a short time and starts as soon as a potential interferer enters the communication range if radar interference and communication signals are subject to the same minimum signal to noise ratio at the RadChat receiver and $\Ptxrad =\Ptxcomm$.  
\end{itemize}

\section{RadChat Performance Evaluation and Results}
\label{sec_results}

The performance of the proposed FMCW-based distributed RadChat protocol is evaluated through Matlab R2017b simulations using the phased array toolbox for a network of vehicles, where a single RadChat unit is mounted on each vehicle with the same FMCW sawtooth waveform parameters in a scenario with a large number of uncoordinated radars. 
Several performance metrics are considered in different dimensions: (i) the probability of R2R interference, (ii) the time it takes for RadChat to minimize interference, (iii) the effect of synchronization errors to RadChat; (iv) the radar jitter; (v) impact of RadChat penetration rate; (vi) effect of the communication parameters.

\subsection{Simulation Parameters}
\label{sec_assumptions}

The main simulation parameters are summarized in Table~\ref{tab:parameters}. 
\begin{table}[t!]
\centering \caption{Simulation parameters.}
%	\footnotesize
 %
\begin{tabular}{lll}
\toprule
 & \textbf{Parameter}  & \textbf{Value} \tabularnewline
\midrule
\multirow{5}{*}{\rotatebox{90}{Radar}}
 & Chirp duration ($T$)  & \SI{20}{\micro\second} \tabularnewline
 & Frame duration ($T_f$)  & \SI{20}{m\second} \tabularnewline
 & Modified duty cycle  ($U'$) & 0.1  \tabularnewline
  %& Slots per frame ($K$)  & \SI{10} \tabularnewline
 & Radar bandwidth ($B_r$)  & \SI{0.96}{\giga\hertz}--\SI{1}{\giga\hertz}  \tabularnewline
  & Bandwidth of interest ($B_\text{max}$)  & \SI{50}{\mega\hertz}  
  \tabularnewline
     & $d_{\text{max}}$ for $B_{\rm{c}}=0$ & 150 m \tabularnewline
        & $v_{\text{max}}$ & 140 km/h \tabularnewline
  & $\Ptxrad$, $\Ptxcomm$   & \SI{11}{\decibel} \tabularnewline
  & SNR  & \SI{10}{\decibel} \tabularnewline
   & Number of chirps per frame ($N$)  & 99 \tabularnewline
    & Carrier frequency ($\fr$) & \SI{77}{\giga\hertz}  \tabularnewline
        & $T_s$ & \SI{0.01}{\micro\second} \tabularnewline
 & Chebyshev low-pass filter order  & 13 \tabularnewline
 & Thermal noise temperature $T_0$ & \SI{290}{\kelvin} \tabularnewline
 & Receiver's noise figure  & \SI{4.5}{\decibel} \tabularnewline
\midrule
\multirow{5}{*}{\rotatebox{90}{Comm.}}
& Communication bandwidth $B_{\rm{c}}$  & \SI{40}{\mega\hertz}  \tabularnewline
%& Communication bandwidth $\Bcomm$  & \SI{10}{\mega\hertz}-\SI{100}{\mega\hertz}  \tabularnewline
 %& $\fr$ \todo{???})  & $77-B_{\rm{cont}}/2$ {\giga\hertz} \tabularnewline
 & Packet size ($N_{\text{pkt}}$) & 4800 Bits \tabularnewline
 & Modulation  & 16-QAM \tabularnewline
 %& Data rate  & X \todo{???}Mbps \tabularnewline
 & MAC  & non-persistent CSMA 
  \tabularnewline
  & SlotTime $\delta$& \SI{10}{\micro\second}
 \tabularnewline
  & Maximum contention window size ($W_0$) & 6
   \tabularnewline
  & Maximum backoff stage ($B$) & 3
  \tabularnewline
\bottomrule
\end{tabular}\label{tab:parameters}
\end{table}
The chirp sequence is designed so as to meet the maximum detectable relative velocity $v_{\text{max}}=\SI{140}{\kilo\meter/\hour}$, the maximum detectable range $d_{\text{max}}=\SI{150}{\meter}$ when $B_{\rm{c}}=0$ (since it increases for RadChat), velocity resolution smaller than \SI{1}{\meter/\second} and range resolution of \SI{15}{\centi\meter}.
Radar front-end-hardware component parameters are taken as in~\cite{refOfMatlab}. The mean value for the radar cross section of a car is taken as \SI{20}{\decibelSM}~\cite{Lee2016,refOfMatlab}. Finally, greatest of cell averaging constant false alarm rate (GoCA-CFAR) thresholding with 50 training cells with 2 guard cells is used for radar detection. We will focus dense networks, so we set $\alpha_d=1$, leading to a vulnerable duration for $B_{\rm{c}}=\SI{40}{\mega\hertz}$ of $| V |=\SI{2.08}{\micro\second}$ \eqref{eq:Vulnerable2}, leading to maximum 7 concurrent radar transmissions per $T_k$, resulting with $M_\text{max}=70$ vehicles supported maximum by RadChat.
A total of 10,000 Monte Carlo simulations are run to obtain interference probability results. The  interference probability was calculated as follows: for each realization and each frame, we declare an occurrence of interference if there was at least 1 interferer present in the vulnerable period in at least 1 chirp within that frame. The interference probability is the number of such occurrences divided by 10,000 and can be visualized as a function of the frame index to show the convergence behavior.

\subsection{Results and Discussion}

\subsubsection{Time to minimize interference}
\begin{figure}
\centering{}\includegraphics[width=1\columnwidth]{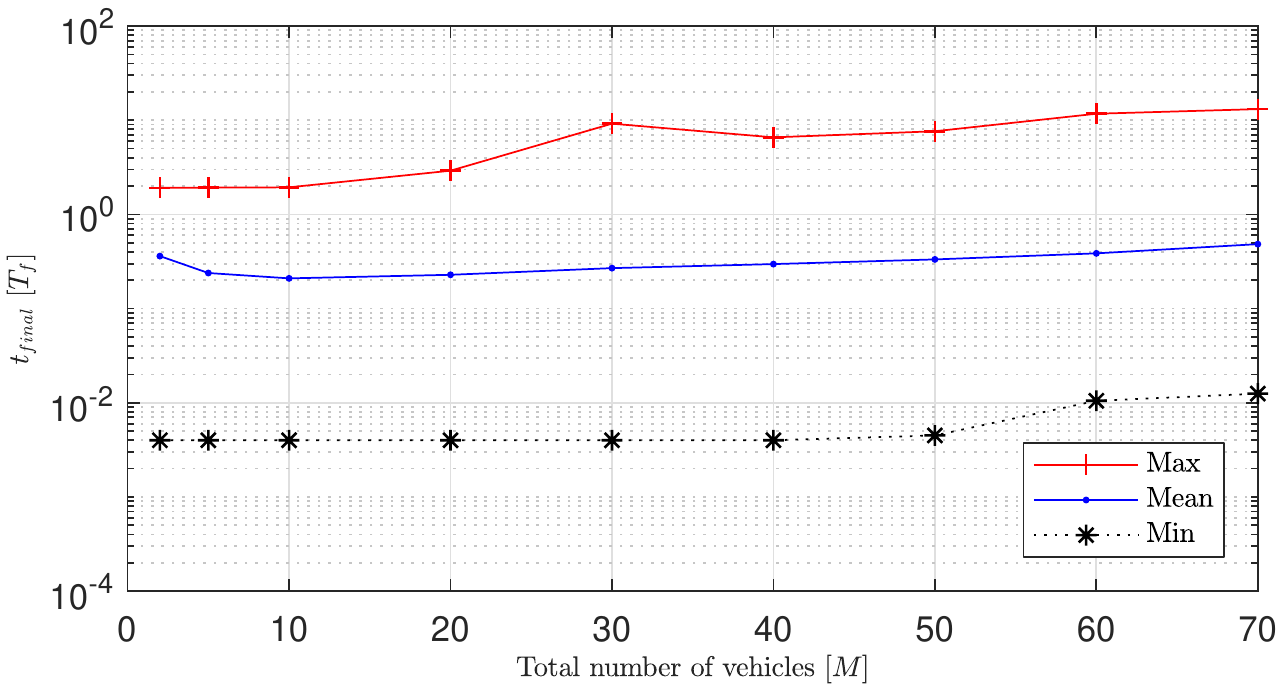}
\caption{The mean, maximum and minimum time to reach negligible R2R interference (no interference among 10,000 simulations) for varying number of vehicles $M$.}
\label{figTfinal}
\end{figure}

Since radars employing RadChat exchange radar starting times, the R2R interference probability vanishes in the steady-state when all facing radars exchange information and select non-overlapping rTDMA slots. 
The time to reach negligible R2R interference \textit{(no interference among 10,000 simulations)} is denoted as $t_{\text{final}}$ and its maximum, mean and minimum value in a network of $M$ vehicles is shown in Fig.~\ref{figTfinal}. RadChat is observed to eliminate interference in less than $T_f/2=$ \SI{10}{\milli\second} on the average, whereas the maximum time to mitigate the interference over 10,000 simulations is less than 13$T_f$, being less than \SI{260}{\milli\second} for $W_0=6$. However, selection of a larger $W_0$ is observed to decrease this maximum time to reach negligible R2R interference to \SI{100}{\milli\second} (Fig.~\ref{fig_CW70All}).

\begin{figure}
\centering{}\includegraphics[width=1\columnwidth]{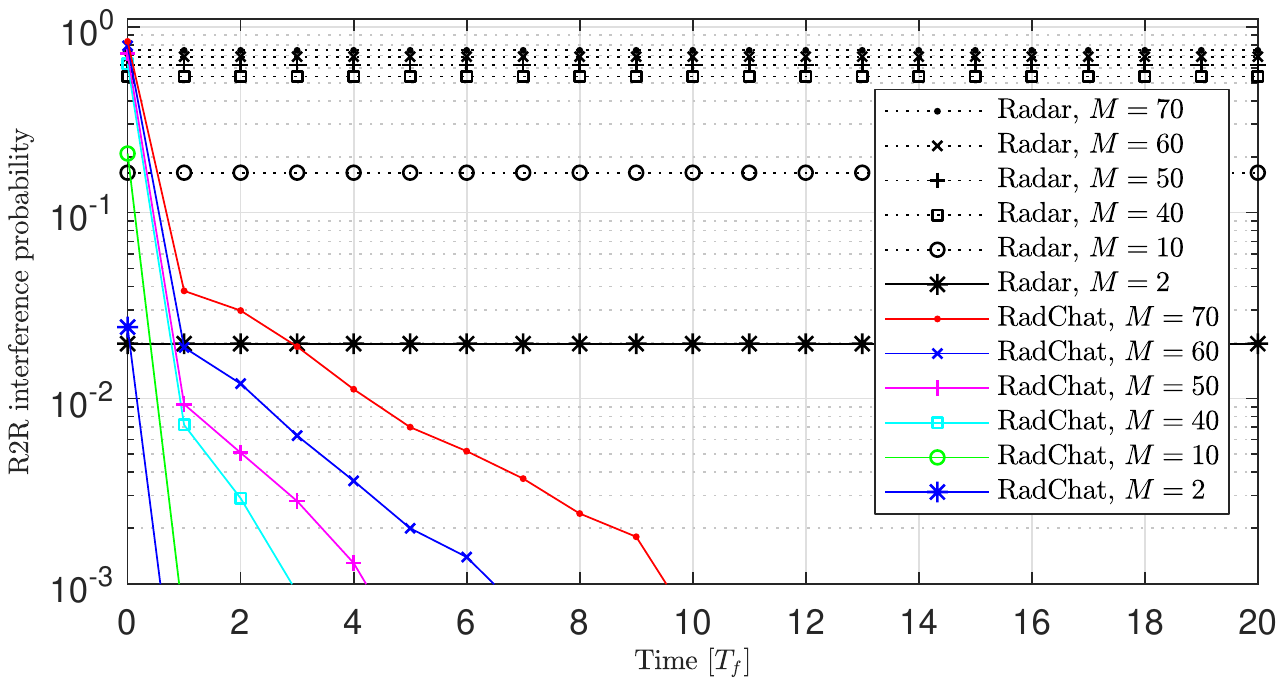}
\caption{Comparison of R2R interference over time for varying $M$ facing radars with RadChat and regular radar.}
\label{figPintRadCom_Tf_M}
\end{figure}

%The R2R interference, averaged over 10000 simulations, which corresponds to $P_{\text{int}}^{(f)}$ of the victim vehicle, is given in Fig.~\ref{figPintRadCom_Sim_Anal} for various $M_{\text{max}}$ compared to radar only case. Since, the total available rTDMA slots are 70 in these simulation settings, the  R2R interference becomes zero in the steady state. After one radar frame duration $T_f$, the mutual interference is decreased by at least $95\%$ with RadChat. Note that, although one frame time is generally enough for a low number of facing vehicles to resolve radar mutual interference totally, it may not be resolved with $10^-4$ probability for two facing radars. It may occur that the communication control packet is not received due to radar transmission (since carrier sensing is inactive during radar transmission) in the first radar frame. Such case is always resolved given that the radar duty cycle is larger than $50\%$.
%\begin{figure}
%\centering{}\includegraphics[width=1\columnwidth]{figures/figPintRadCom_Sim_Anal.pdf}
%\caption{Comparison of the R2R interference for varying $M_{\text{max}}$ facing radars with radar only %case.}
%\label{figPintRadCom_Sim_Anal}
%\end{figure}
Fig.~\ref{figPintRadCom_Tf_M} shows how the R2R interference decreases over time with RadChat. It is observed that the R2R interference decreases sharply after one frame duration (by more than a factor of 25) and below $10^{-3}$ within less than 10 frames (200 ms) for the case with 70 interfering radars. 

%\subsection{Time to reach steady state: $t_{final}$}
%For a mobile vehicular network with no additions and deletions, the time to reach the zero R2R interference, i.e. steady state, is denoted as $t_{\text{final}}$ and shown in Fig..~\ref{figTfinal}. 
%\begin{figure}
%\centering{}\includegraphics[width=1\columnwidth]{figures/figTfinal.pdf}
%\caption{The mean, max and min time to reach zero R2R interference for varying $M_{\text{max}}$.}
%\label{figTfinal}
%\end{figure}

\subsubsection{Effect of synchronization error $\varepsilon_{\text{sync}}$}
\begin{figure}
\centering{}\includegraphics[width=1\columnwidth]{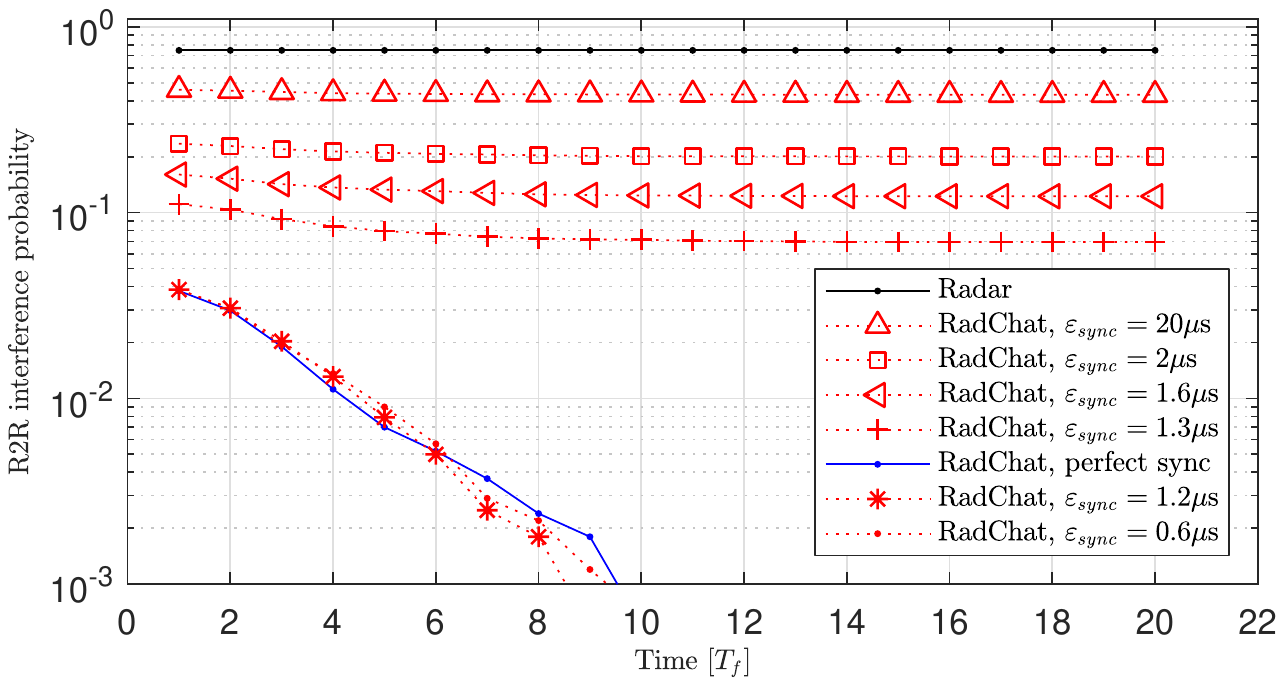}
\caption{Comparison of the R2R interference probability for the regular radar with RadChat with zero and varying synchronization error, for $M=70$.}
\label{figPintRadCom_Tf}
\end{figure}

The R2R interference over time in a network of $M_=70$ vehicles is compared with various synchronization errors in Fig.~\ref{figPintRadCom_Tf}. A uniformly distributed synchronization error with maximum values $\varepsilon_{\text{sync}}=\pm\{0.6,1.2,1.3,1.6,2,20\}\ \SI{}{\micro\second}$ is assumed, where each node is assumed to retain the same synchronization error during the simulation. Our simulations are based on a discrete time resolution of $\SI{0.2}{\micro\second}$, which led to $V = \SI{2.4}{\micro\second}$ after rounding. Hence, the impact of synchronization errors for $\varepsilon_{\text{sync}} \leq V/2 = \SI{1.2}{\micro\second}$ is expected to be minor for RadChat, performing almost the same as with perfect synchronization. For $\varepsilon > \SI{1.2}{\micro\second}$, the possibility of overlap of radar chirp sequences leads to a high floor of the interference probability. 

%\CA{A synchronization error $\varepsilon$ less than and equal to $\pm1.2\mu$s is observed not to affect the RadChat performance significantly. However for if $\varepsilon \geq \SI{1.3}{\micro\second}$s, RadChat cannot schedule radars to non-overlapping rTDMA slots and cannot eliminate mutual interference as time proceeds. The resulting R2R interference probability with RadChat is smaller than the radar only case since RadChat schedules radars to different time slots $T_k$ and since we have 10 different such time slots, the victim radar can at most interfere with radars scheduled inside its own and former time slot.}

%\CA{The cut-off value for the synchronization error is equal to half of the vulnerable period. The vulnerable period for $B_{\rm{c}}=40$ MHz is $2.0833\mu$s. Due to the discretization of time during the simulation settings by $0.2\mu$s intervals, the duration between two rTDMA discrete slots is equal to $2.4\mu$s. Hence, a synchronization error of $\varepsilon \leq 1.2 \mu$ in the discretized time frame results with no overlapping radar $B_\text{max}$s. With a perfect synchronization radars could be packed by $\vert V\vert/2$ duration in between, but by placing each radar transmission in the middle of a vulnerable period, we leave a guard time for synchronization errors of $\vert V\vert/2$. Any synchronization error above this, leads to loss of RadChat benefits.}

\subsubsection{Radar Jitter}
The periodicity of radar is observed to be distorted at most by one $T_f$ during the initialization stage and the radar experiences no jitter afterwards (result not shown). 
%The proposed RadChat scheme reschedules radar transmissions so that facing radars do not send in overlapping vulnerable periods. This mechanism introduces delays for the radar. The instantaneous jitter in radar transmission times is shown in Fig.~\ref{figDelay} for $M=70$ over time for $W_0=6$. 
%Given that the radar frame duration is 20 ms, it is observed that the periodicity of radar is distorted at most by one $T_f$ and the radar experiences no jitter after that. 
%\begin{figure}
%\centering{}\includegraphics[width=1\columnwidth]{figures/figDelay.pdf}
%\caption{Instantaneous radar jitter for each of radars with a total of $M=70$ facing radars ($W_0=6$, $B_{\rm{c}}=\SI{40}{\mega\hertz}$).}
%\label{figDelay}
%\end{figure}

\subsubsection{RadChat Deployment}

\begin{figure}
\centering{}\includegraphics[width=1\columnwidth]{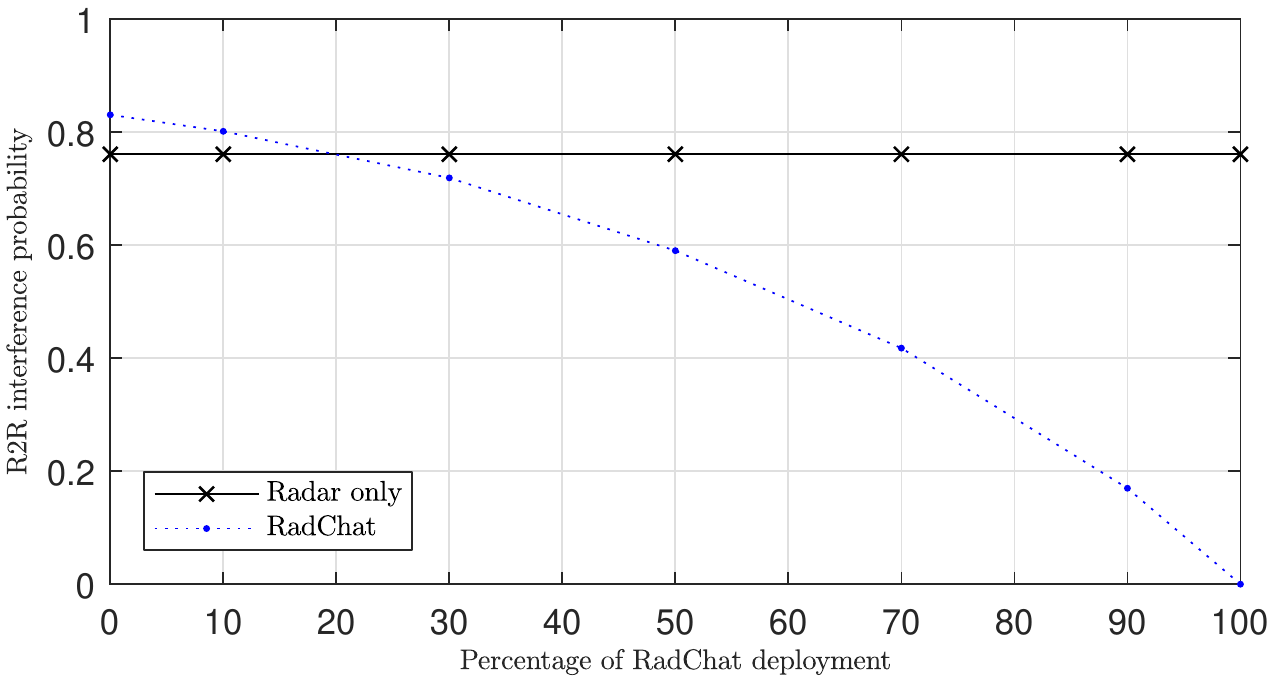}
\caption{R2R interference probability versus percentage of RadChat deployment for a network of $M=70$ facing radars.}
\label{figPintDeaf}
\end{figure}

We investigate the R2R interference probability in heterogeneous network setting where not all nodes are equipped with RadChat. The R2R interference experienced by a vehicle is compared for a network of $M=70$ vehicles with changing percentages of RadChat equipped vehicles in Fig.~\ref{figPintDeaf}, where $B_{\rm{r}}=\SI{1}{\giga\hertz}$ for radar only case and $B_{\rm{r}}=\SI{0.96}{\giga\hertz}$ for RadChat. 
The results show that $100\%$ deployment of RadChat results with almost total elimination of R2R interference, though these benefits diminish very quickly with reduced deployment. When no vehicles are equipped with RadChat, the reduction of the available radar bandwidth $B_{\rm{r}}$ increases R2R interference due to (\ref{eq:chirpInt}). 

%whereas $50\%$ deployment results with $28\%$ decrease and less than about $20\%$ deployment of RadChat results with worse R2R interference. Hence, the proposed scheme is able to reduce R2R interference considerably if employed by almost all vehicles, otherwise the benefits diminish very quickly. 

\begin{figure}
\centering{}\includegraphics[width=1\columnwidth]{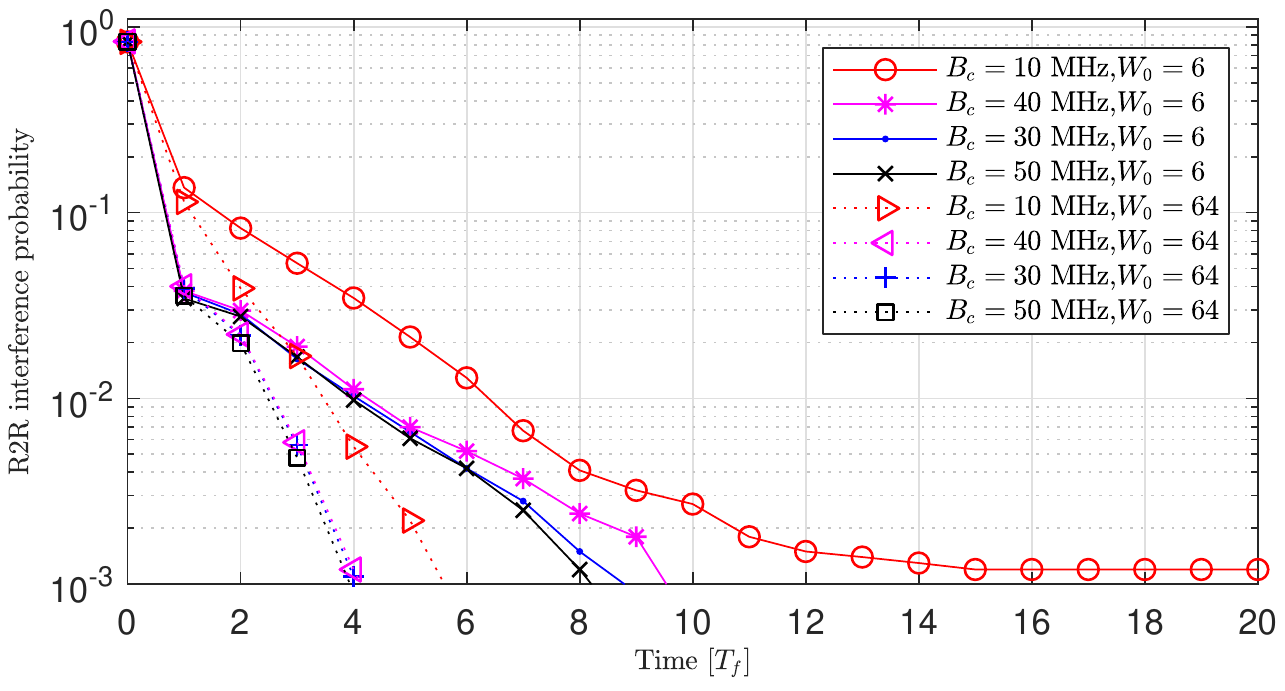}
\caption{R2R interference probability versus time for changing $B_{\rm{c}}$ and $W_0$ for $M=70$.}
\label{figBc1}
\end{figure}

\begin{figure}
\centering{}\includegraphics[width=1\columnwidth]{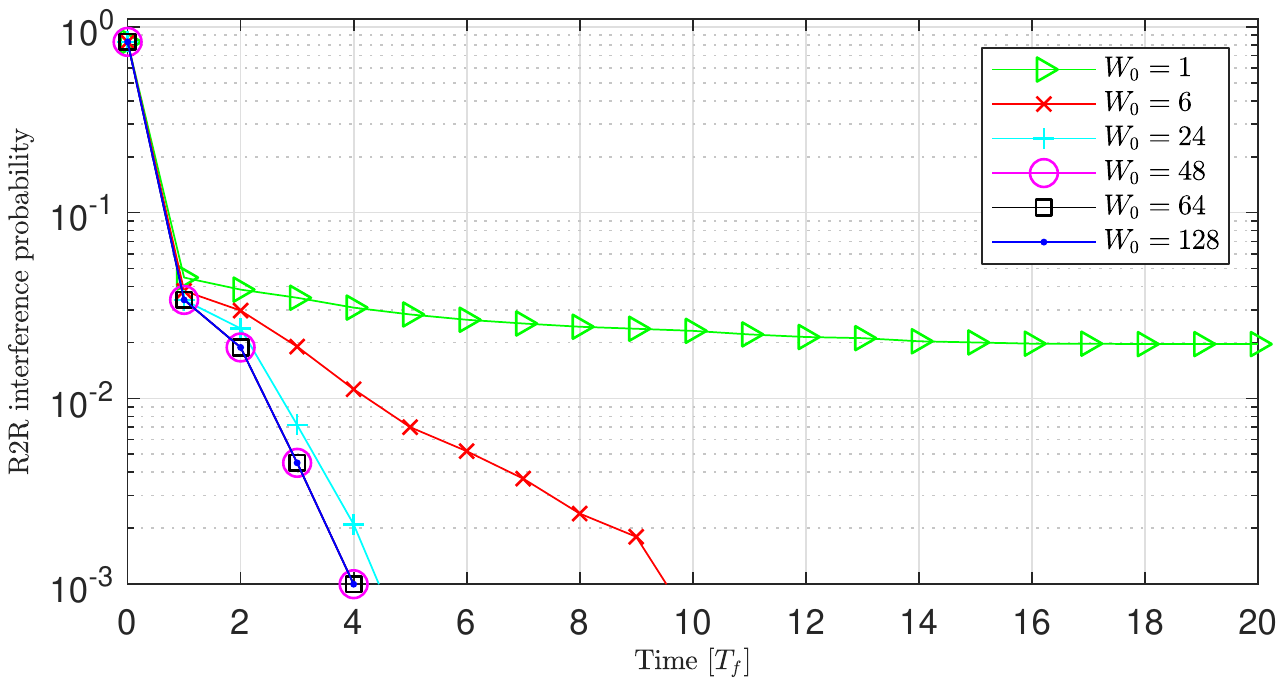}
\caption{R2R interference probability versus time for varying $W_0$ for $M=70$.}
\label{fig_CW70}
\end{figure}

\begin{figure}
\centering{}\includegraphics[width=1\columnwidth]{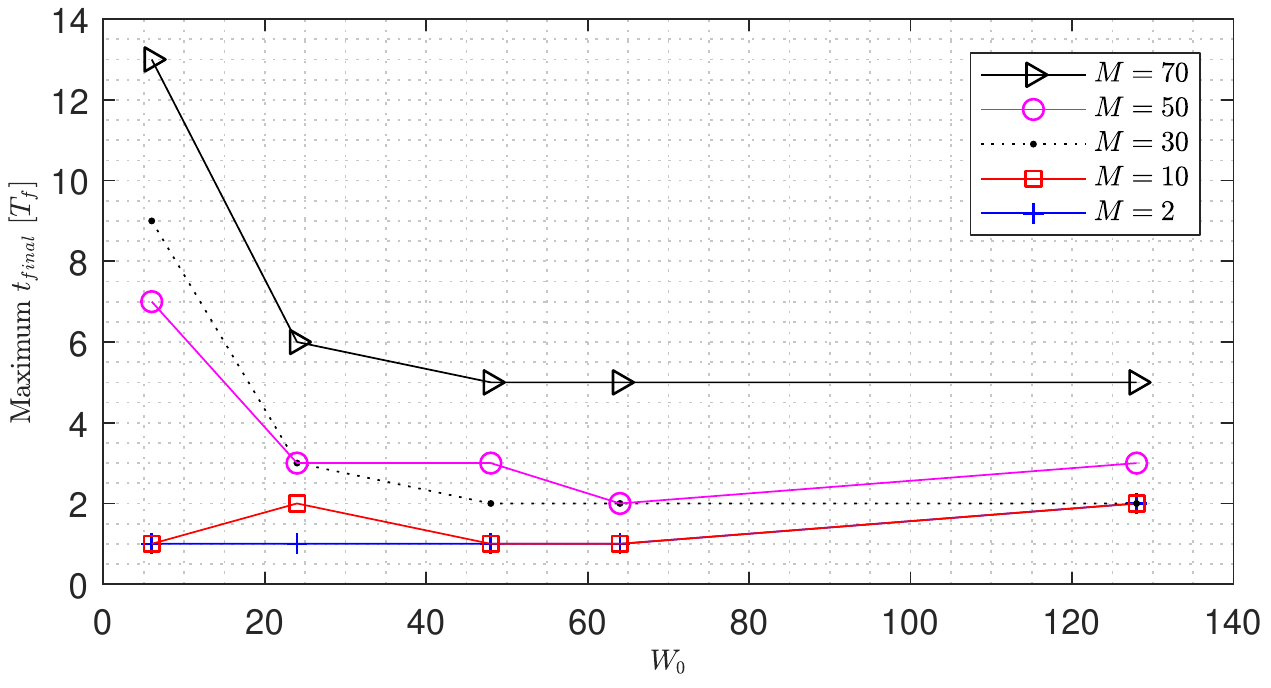}
\caption{The maximum time to reach negligible R2R interference (no interference among 10,000 simulations) for varying $M$ and $W_0$.}
\label{fig_CW70All}
\end{figure}

\subsubsection{Effect of Communication Parameters}
The effect of $B_{\rm{c}}$ on R2R interference is investigated for $W_0=\{6,64\}$ in Fig.~\ref{figBc1}. In order for RadChat to converge, a large enough bandwidth must be assigned to the control channel. With a larger bandwidth, convergence is slightly faster, while with a small bandwidth, there is a floor in the interference probability. Note that allocation of a portion of bandwidth to communication comes with a cost of degradation in the radar range estimation performance. However, the radar performance degradation is negligible for automotive applications for the considered communication bandwidths (0.63 cm radar range resolution reduction and 1.64 cm range estimation error increase~\cite{CananRadConf}). In Fig.~\ref{fig_CW70}, the maximum value of $t_{\text{final}}$ is observed to be highly affected by maximum contention window size $W_0$ for $M=70$. Maximum value of $t_\text{final}$ is observed to decrease from 10$T_f$ to 4$T_f$ with a change of maximum contention window size from 6 to 64. Both the communication bandwidth and the contention window size affect the convergence time considerably due to the CSMA-based contention scheme.

The maximum value of $t_{\text{final}}$ for varying $M$ and $W_0$ is given in Fig.\ref{fig_CW70All}. It is observed that there is an optimum $W_0$ for reaching the steady-state as quickly as possible in the worst case, which depends on $M$. RadChat converges in at most $t_{\text{final}}=1 T_f$ for $M\leq 10$ with $W_0=\{48,64\}$, whereas $t_{\text{final}}=5 T_f$ for $M=70$ with $W_0\geq 48$. This indicates that the best choice for the maximum contention window size for the given parameters in Table \ref{tab:parameters} is $W_0=64$, which ensures that RadChat is able to reduce R2R interference below $10^{-3}$ in \SI{80}{\milli\second} and a reduction of an order of magnitude is achieved almost in one radar frame time (not shown in results) for a newly formed 70-vehicle VANET. This duration is expected to be shorter when multiple radars are mounted per vehicle and VANET connectivity changes slowly.

\section{Conclusion}

Based on interference analyses for spectrum sharing of automotive radar and vehicular communications, we propose guidelines for mitigation of interference and designed the radar and communication cooperation system RadChat for FMCW-based automotive radar interference mitigation. RadChat builds upon the same hardware for radar and communications and a MAC, which is a combination of FDM, TDMA for radar, and CSMA for communication. RadChat exploits the low utilization of time and frequency of a typical radar with the limited impact of a small bandwidth loss on the radar performance. Extensive network simulations show that automotive radar interference probability is reduced significantly, by about one order of magnitude every radar frame time in dense VANETs. RadChat is expected to mitigate R2R interference even in sparse networks by adaptation of the vulnerable period in combination with fewer interfering vehicles.   With our proposed approach, we are able to mitigate interference by shifting radar transmissions in time with higher penetration rate. %R2R interference mitigation performance is reported  %based on network simulations. 
Future work will consider larger-scale scenarios for heterogeneous FMCW radars with different bandwidths and chirp parameters, as well as the additional use of RadChat for inter-vehicle data communication. %\textcolor{blue}{Furthermore, extensions to RadChat in order to include data communications additional to interference mitigation functionality will be developed}. 

\appendices
 \section{Vulnerable period for R2R interference}
The transmission by Vehicle 2, which starts at time $\tau$ is received by Vehicle 1 in Fig.\ref{fig_scenarioIntR2R} at time $t'$ and is equivalent to a chirp reception starting at $t'-\tau_D$ without any Doppler shift. $\tau_D$ is the perceived Doppler time delay and is obtained as follows after applying the triangle proportionality theorem to one FMCW chirp in Fig.~\ref{fig_chirpJournal}  
\begin{align}
    \tau_D & = T f_D/B_\text{r}={T v \fr}/{(B_\text{r} c)} \\
     & \in [-T |v| \fr/{(B_\text{r} c)},+T |v| \fr/{(B_\text{r} c)}] \\
     & \subset [- {T |v_{\text{max}}| \fr}/{(B_\text{r} c)},+{T |v_{\text{max}}| \fr}/{(B_\text{r} c)}]\\
     & \approx [-1/(4B_\text{r}),+1/(4B_\text{r})]
\end{align}
where we have made use of the fact that the maximum radar detectable relative velocity is given by $v_{\text{max}}=c/(4\fr T)$~\cite{skolnik}, and that vehicles may approach or recede. 

R2R interference
at the first chirp occurs when $t'-\tau_D\in [0,T_\text{max}]$, i.e.. when the first red chirps falls inside the green-coloured region in Fig.~\ref{fig_chirpJournal}, where $T_\text{max} = T B_\text{max}/B_\text{r}$ corresponds to the maximum delay for detectable radar targets. Considering all possible distances, ${d}/{c} \in [0,\alpha_d T_\text{max}]$, then vulnerable period is given by 
\begin{align}
   V = \left [-\alpha_d T_\text{max} -{1}/{(4B_\text{r})}, T_\text{max}+{1}/{(4B_\text{r})} \right] \label{eq:VulnerableDuration}
\end{align}
In practice, $B_\text{r} \gg 1/T_\text{max}$ so this term may be ignored. Hence, the vulnerable period is approximately $V=\left [-\alpha_d T_\text{max},T_\text{max}\right]$, assuming a radar that can sample in-phase and quadrature samples and has perfect low-pass filtering. 

\bibliographystyle{IEEEtran}
\bibliography{IEEEabrv,./bibliography}

\begin{IEEEbiography}[{\includegraphics[width=1in,height=1.25in,clip,keepaspectratio]{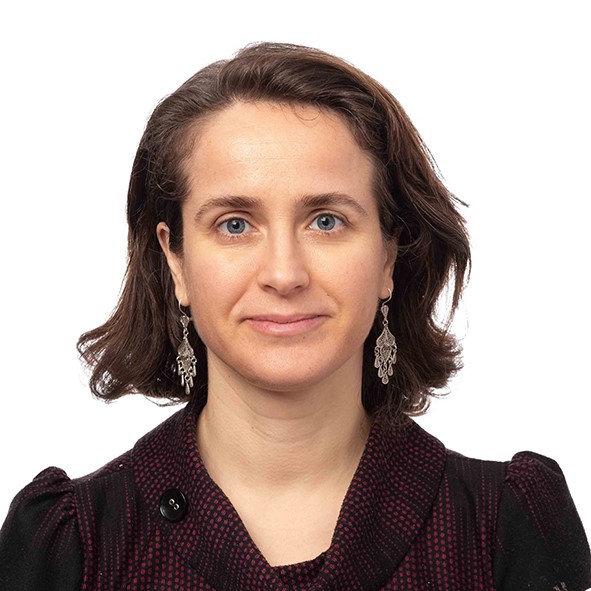}}]{Canan Aydogdu} is a Marie Sklodowska-Curie Fellow at Chalmers, working on cross-layer multi-hop sensing systems for the future energy-efficient self-driving car networks. Formerly, she was an associate professor at Izmir Institute of Technology, Turkey. She received the Ph.D. degree from Bilkent University, Turkey, in 2011.
\end{IEEEbiography}
%\vspace{-0.45in}

\begin{IEEEbiography}[{\includegraphics[width=1in,height=1.25in,clip,keepaspectratio]{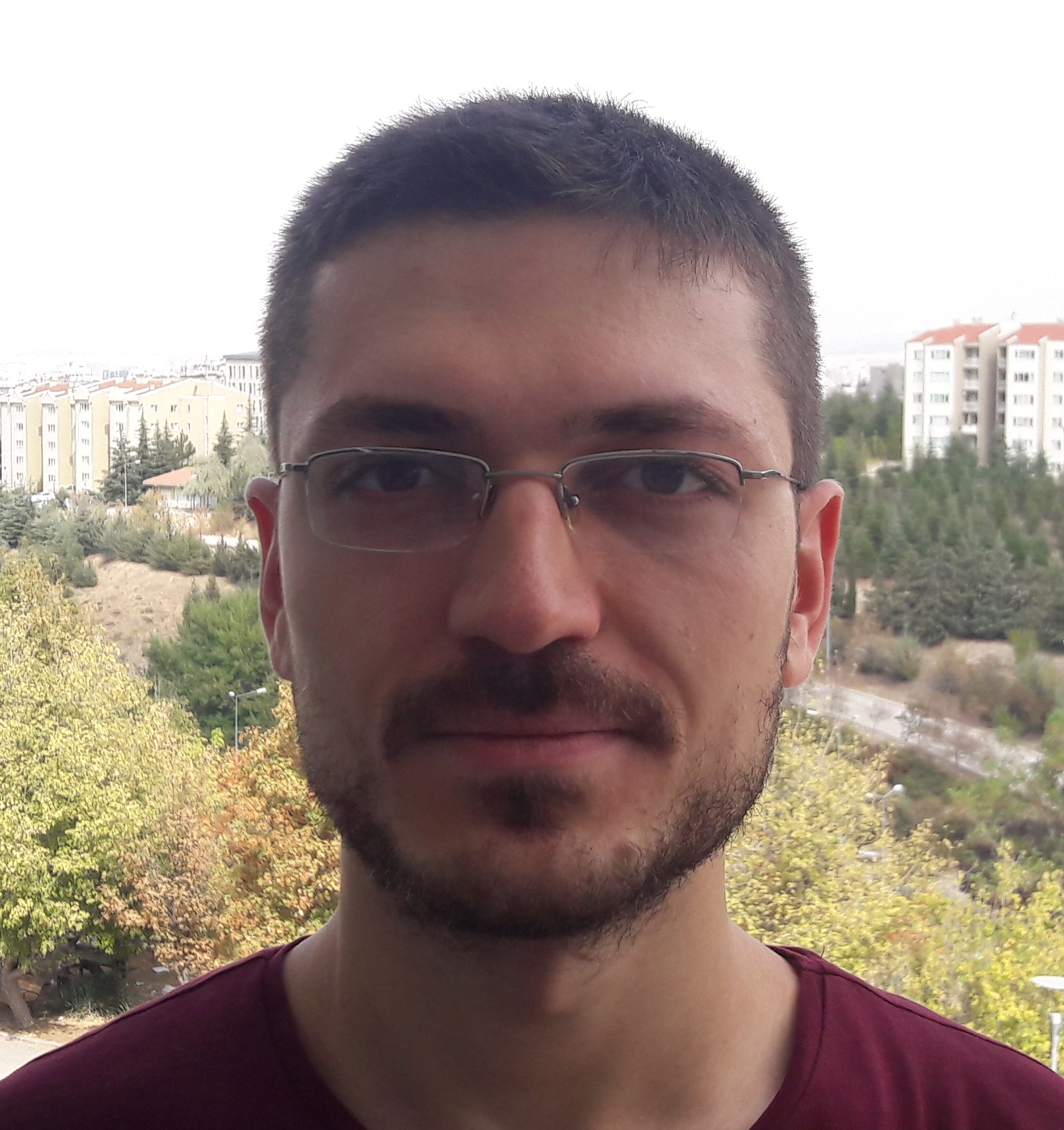}}]{Musa Furkan Keskin} is a postdoctoral researcher at Chalmers University of Technology. He obtained the B.S., M.S., and Ph.D degrees from the Department of Electrical and Electronics Engineering, Bilkent University, Ankara, Turkey, in 2010, 2012, and 2018, respectively. His current research interests include intelligent transportation systems, radar signal processing and radar communications coexistence.
\end{IEEEbiography}
%\vspace{-0.45in}

\begin{IEEEbiography}[{\includegraphics[width=1in,height=1.25in,clip,keepaspectratio]{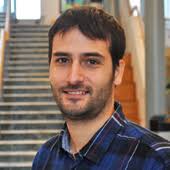}}]{Nil~Garcia}
        (S'14--M'16) received the Telecommunications Engineer degree from the Polytechnic University of Catalonia (UPC), Barcelona, Spain, in 2008; and the double Ph.D.\ degree in electrical engineering from the New Jersey Institute of Technology, Newark, NJ, USA,  and from the National Polytechnic Institute of Toulouse, Toulouse, France, in 2015.
		
		He is currently a postdoctoral researcher of Communication Systems with the Department of Signals and Systems at Chalmers University of Technology, Sweden. In 2009, he worked was an engineer in the Centre National d'\'{E}tudes Spatiales (CNES). In 2008 and 2009, he had internships in CNES and NASA. Hi research interests are in the areas of localization, intelligent transportation systems and 5G.
\end{IEEEbiography}

%\vspace{-0.45in}

\begin{IEEEbiography}[{\includegraphics[width=1in,height=1.25in,clip,keepaspectratio]{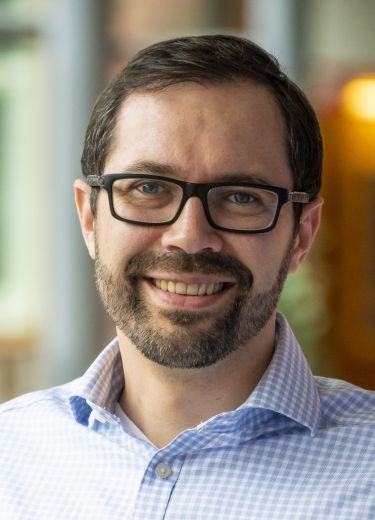}}]{Henk~Wymeersch}
(S'01, M'05) obtained the Ph.D. degree in Electrical Engineering/Applied Sciences in 2005 from Ghent University, Belgium. He is currently a Professor of Communication Systems with the Department of Electrical Engineering at Chalmers University of Technology, Sweden and Distinguished Research Associate with Eindhoven University of Technology, the Netherlands. Prior to joining Chalmers, he was a postdoctoral researcher with the Laboratory for Information and Decision Systems at the Massachusetts Institute of Technology. His current research interests include cooperative systems and intelligent transportation. \end{IEEEbiography}
%\vspace{-0.45in}

\begin{IEEEbiography}[{\includegraphics[width=1in,height=1.25in,clip,keepaspectratio]{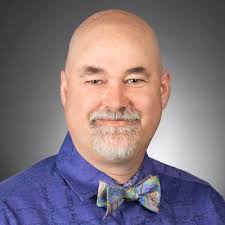}}]{Daniel~W.~Bliss}
Daniel W. Bliss is an Associate Professor in the School of Electrical, Computer, and Energy Engineering at ASU and a Fellow of the IEEE.  He is also the Director of Arizona State University's Center for Wireless Information Systems and Computational Architectures.  Dan received his Ph.D. and M.S. in Physics from the University of California at San Diego (1997 and 1995), and his B.S. in Electrical Engineering from Arizona State University (1989). His current research focuses on advanced systems in the areas of communications, radar, precision positioning, and medical monitoring.  Dan has been the principal investigator on numerous programs including sponsored programs with DARPA, ONR, Google, Airbus, and others. He is responsible for foundational work in electronic protection, adaptive multiple-input multiple-output (MIMO) communications, MIMO radar, distributed-coherent systems, and RF convergence. Before moving to ASU, Dan was a Senior Member of the Technical Staff at MIT Lincoln Laboratory (1997-2012). Between his undergraduate and graduate degrees, Dan was employed by General Dynamics (1989-1993), where he designed avionics for the Atlas-Centaur launch vehicle, and  performed magnetic field optimization for high-energy particle-accelerator superconducting magnets.  His doctoral work (1993-1997) was in the area of high-energy particle physics and lattice-gauge-theory calculations. Dan is a member of the IEEE AES Radar Systems Panel and is a member of the IEEE Signal Processing Magazine editorial board.
\end{IEEEbiography}
\vspace{-0.45in}

\end{document}